\documentclass[aps,prd,superscriptaddress,nofootinbib]{revtex4}
\usepackage[a4paper, total={6.0in, 9in}]{geometry}
\usepackage{amsfonts,amssymb,latexsym,amsmath,mathrsfs} 
\usepackage{bm}
\usepackage{hhline}
\usepackage{longtable}
\usepackage{mathtools} 
\usepackage{array} 
\usepackage{color}
\usepackage{ifthen}
\usepackage{epsfig}
\usepackage{multirow}
\usepackage{enumerate}
\usepackage{dsfont}
\usepackage{psfrag}
\usepackage{float}
\usepackage{appendix}
\usepackage{slashed}
\usepackage{xcolor,booktabs}
\graphicspath{{Figures/}} 
\usepackage{hyperref}
\hypersetup{colorlinks,citecolor=blue,anchorcolor=red,menucolor=red, linkcolor=red,filecolor=red,runcolor=red,urlcolor=blue}

\begin{document}
	
\title{Radiative decay and axial-vector decay behaviors of octet pentaquark states}
\author{Ya-Ding Lei}
\email{yadinglei@stumail.nwu.edu.cn}
\affiliation{School of Physics, Northwest University, Xian 710127, China}
\author{Hao-Song Li}
\thanks{Corresponding author}
\email{haosongli@nwu.edu.cn}
\affiliation{School of Physics, Northwest University, Xian 710127, China}
\affiliation{Institute of Modern Physics, Northwest University, Xian 710127, China}
\affiliation{Shaanxi Key Laboratory for Theoretical Physics Frontiers, Xian 710127, China}
\affiliation{Peng Huanwu Center for Fundamental Theory, Xian 710127, China}

\begin{abstract}
	In this work, we systematically calculate transition magnetic moments, radiative decay widths, and axial-vector coupling constants of octet hidden-charm molecular pentaquark states with different flavor representations in constituent quark model. We discuss the relations between transition magnetic moments and decay widths for pentaquark states. For octet pentaquark states with the $8_{1f}$ and $8_{2f}$ flavor representations, decay widths of the processes $P_{\psi}|\frac{3}{2}^-\rangle_{(\frac{1}{2}^+\otimes1^-)}\to P_{\psi}|\frac{1}{2}^-\rangle_{(\frac{1}{2}^+\otimes0^-)}\gamma$ and $P_{\psi}|\frac{1}{2}^-\rangle_{(\frac{1}{2}^+\otimes1^-)}\to P_{\psi}|\frac{1}{2}^-\rangle_{(\frac{1}{2}^+\otimes0^-)}\gamma$ are quite close, decay widths of the $P_{\psi}|\frac{3}{2}^-\rangle_{(\frac{1}{2}^+\otimes1^-)}\to P_{\psi}|\frac{1}{2}^-\rangle_{(\frac{1}{2}^+\otimes1^-)}\gamma$ process are close to zero, and we notice that the axial-vector coupling constants of the pentaquark states are generally smaller than that of the nucleon.
\end{abstract}
\maketitle

\section{Introduction}\label{sec1}
In 2003, X(3872) was observed by Belle Collaboration, which led to an explosive exploration of exotic hadronic states\cite{Belle:2003nnu}. In the past decades, a large number of exotic hadronic states were experimentally observed, which stimulated many discussions\cite{Chen:2016qju,Hosaka:2016pey,Lebed:2016hpi,Meng:2022ozq,Chen:2022asf,Ali:2018ifm,Ali:2018xfq,Francis:2018jyb,Fontoura:2019opw,Junnarkar:2018twb}. In 2015, the pentaquark states $P_{\psi}(4380)^+$ and $P_{\psi}(4450)^+$ were observed in the $\Lambda_b^0\to J/\psi p K^-$ decays by the LHCb Collaboration, which are close to
the $\Sigma_c^*\bar{D}$ and $\Sigma_c\bar{D}^*$ thresholds\cite{LHCb:2015yax}. It is the first time that the pentaquark state was observed experimentally. In 2019, the LHCb
Collaboration observed that the $P_{\psi}(4450)^+$ consists of two narrow overlapping peaks, $P_{\psi}(4440)^+$ and $P_{\psi}(4457)^+$, and reported a new state $P_{\psi}(4312)^+$ near $\Sigma_c\bar{D}$ threshold\cite{LHCb:2019kea}. Subsequently LHCb
Collaboration announced the existence of a series of pentaquark states such as $P_{\psi}(4459)^+$\cite{LHCb:2020jpq}, $P_{\psi}(4337)^+$\cite{LHCb:2021chn} and $P_{\psi}(4338)^+$\cite{LHCb:2022ogu}, and these $P_{\psi}$ states are considered to be hidden-charm pentaquark state candidates. The observation of pentaquark states aroused a great deal of interest among researchers, and theorists engaged in extensive discussions about the nature of pentaquark states\cite{Wang:2015ava,Chen:2015sxa,Lu:2016roh,Wang:2019ato,He:2019rva,Pimikov:2019dyr,Pan:2019skd,Ali:2019clg,Voloshin:2019aut,Mutuk:2019snd,Weng:2019ynv,Du:2021fmf,Zhu:2021lhd,Hu:2021nvs,Wang:2021hql,Chen:2021tip,Zhu:2022wpi,Nakamura:2022gtu,Ozdem:2022kei,Wang:2022mxy}.

Exploring exotic hadronic states can deepen our understanding of the nonperturbative behavior of QCD in the low-energy regime, and help us to understand more about the inner structure of hadrons. The axial-vector coupling constant is another important physical quantity which can help us analyze the electroweak and strong interactions in the Standard Model, and it is also an indicator of non-perturbative QCD chiral symmetry breaking. The inner structure of pentaquark states has been investigated by various methods, such as quark model\cite{Ortega:2022uyu,Deng:2022vkv,Hiyama:2018ukv}, QCD sum rules\cite{Zhang:2019xtu,Chen:2016otp,Chen:2019bip}, chromomagnetic interaction (CMI) model\cite{Li:2023aui,Cheng:2019obk,An:2020vku}, and one-boson-exchange (OBE) model\cite{Chen:2019asm,Chen:2017jjn,Wang:2019nwt}. The masses and decay behavior of pentaquark states are studied in some pictures including molecular
picture\cite{He:2019ify,Burns:2015dwa,Huang:2015uda,Chen:2016heh}, diquark–diquark–antiquark picture\cite{Li:2015gta,Wang:2015epa,Shi:2021wyt}, diquark–triquark picture\cite{Zhu:2015bba,Guo:2023fih,Giron:2021sla} and compact pentaquark picture\cite{Cheng:2021gca,Stancu:2021rro,Kuang:2020bnk}. In Ref. \cite{Wang:2022tib}, the authors systematically investigated magnetic moments, transition magnetic moments and radiative decay behaviors of the $S$-wave isoscalar $\Xi_c^{(\prime)}\bar{D}^{(\prime)}$ molecular pentaquark states in constituent quark model. 
In Ref. \cite{Meng:2019ilv}, the authors calculated the effective potentials of the $\Sigma_c\bar{D}^*$ systems with the heavy hadron chiral perturbation theory, and employed quark model with heavy quark spin symmetry to get some relations between different systems. In Ref. \cite{Duan:2024uuf}, the authors studied the mass spectrum of doubly charmed pentaquark states in the $\Lambda_c^{(*)}D^{(*)}$ and $\Sigma_c^{(*)}D^{(*)}$ channels with $J^P=\frac{1}{2}^{\pm}, \frac{3}{2}^{\pm}$ and $\frac{5}{2}^{\pm}$ within the framework of QCD sum rules. In Ref. \cite{Wang:2023aob}, the authors predicted the mass spectrum of $\Omega_c^{(*)}D_s^{(*)}$-type doubly charmed molecular pentaquark candidates with OBE model considering both the $S$-$D$ wave mixing effect
and the coupled channel effect.

Previously, we systematically studied magnetic moments and axial charges of the octet hidden-charm pentaquark states in Ref. \cite{Li:2024wxr} with constituent quark model. We noticed that the axial charges of octet hidden-charm pentaquark states are generally smaller than that of the nucleon. In this work, we continue to explore further the radiative decay and axial-vector decay behaviors of the octet hidden-charm pentaquark. The properties of these states can provide more information about the inner structure of the hidden-charm pentaquark states, which are also helpful for the studies of chiral effective theory.

This paper is organized as follows. In Sec. \ref{sec2}, we construct the wave functions of octet hidden-charm pentaquark states. In Sec. \ref{sec3}, we calculate the transition magnetic moments and radiative decay widths of the octet hidden-charm pentaquark states, and we conclude their relations. In Sec. \ref{sec4}, we construct the Lagrangians of the axial-vector decays and calculate the axial-vector coupling constants of the pentaquark states. Finally, we provide a brief summary in Sec. \ref{sec4}.

\section{Wave functions}\label{sec2}
The total wave function of hadronic state can be expressed as
\begin{eqnarray}
	\Psi=\phi_{\rm flavor}\chi_{\rm spin}\xi_{\rm color}\eta_{\rm space},
\end{eqnarray}
where $\phi_{\rm flavor}$, $\chi_{\rm spin}$, $\xi_{\rm color}$, $\eta_{\rm space}$ are flavor wave function, spin wave function, color wave function and spatial wave function respectively. The total wave function of the pentaquark states requires the fermion exchange symmetry to be satisfied, thus the total wave function is required to be antisymmetric. For the $S$-wave pentaquark states we discussed in this work, the color wavefunction is antisymmetric and the spatial wavefunction is symmetric, so the spin-flavor wavefunction $\phi_{\rm flavor}\chi_{\rm spin}$ should be symmetric.

In the molecular model $(q_1q_2Q)(\bar{Q}q_3)$, we construct the wave functions of the pentaquark with $SU(3)$ symmetry. Here, $q$ and $Q$ denote the light quark and the heavy quark, respectively. In the flavor space, the three light quarks $q_1q_2q_3$ form the flavor representations
\begin{eqnarray}
	(3\otimes3)\otimes3=(6\oplus\bar{3})\otimes3=10_f\oplus8_{1f}\oplus8_{2f}\oplus1_f,
\end{eqnarray}
the flavor wave functions of the pentaquark states can be obtained by adding the heavy quark $c$ and the antiquark $\bar{c}$ to the flavor wave functions of the three light quarks. In the molecular model, the pentaquark states are composed of charmed baryons and anti-charmed mesons. Based on the flavor symmetry of light diquarks, the charmed baryons can be divided into $\bar{3}_f$ and $6_f$ flavor representations, $\bar{3}_f$ represents the flavor antisymmetry of light diquark, and $6_f$ represents the flavor symmetric of light diquark, we list the quark constituents of the charmed baryons and the anti-charmed mesons with different flavor representations in Table \ref{TableQC}. For example, $\Xi^{\prime0}_c$ represents the $S$-wave charmed baryon in the $6_f$ flavor representation, and $\Xi^0_c$ represents the $S$-wave charmed baryon in the $\bar{3}_f$ flavor representation. The charmed baryons and anti-charmed mesons can form the ${8}_{1f}$ and ${8}_{2f}$ flavor representations, and we list the flavor wave functions of the octet hidden-charm molecular pentaquark states in Table \ref{TableFW1}.

\renewcommand\tabcolsep{0.65cm}
\renewcommand{\arraystretch}{1.60}
\begin{table*}[!htbp]
	\caption{The quark constituents of charmed baryons and anti-charmed mesons.}
	\label{TableQC}
	\begin{tabular}{c|c|c|c}
		\toprule[1.0pt]
		\toprule[1.0pt]
		Hardons&Quark constituents&Hardons&Quark constituents\\
		\hline
		$\Sigma_c^{++}$&$uuc$&$\Xi_c^+$&$\sqrt{\frac{1}{2}}(usc-suc)$\\
		\hline
		$\Sigma_c^+$&$\sqrt{\frac{1}{2}}(udc+duc)$&$\Xi_c^0$&$\sqrt{\frac{1}{2}}(dsc-sdc)$\\
		\hline
		$\Sigma_c^0$&$ddc$&$\Lambda_c^+$&$\sqrt{\frac{1}{2}}(udc-duc)$\\
		\hline
		$\Xi_c^{\prime+}$&$\sqrt{\frac{1}{2}}(usc+suc)$&$\bar{D}^{(*)0}$&$\bar{c}u$\\
		\hline
		$\Xi_c^{\prime0}$&$\sqrt{\frac{1}{2}}(dsc+sdc)$&$D^{(*)-}$&$\bar{c}d$\\
		\hline
		$\Omega_c^0$&$ssc$&$D_s^{(*)-}$&$\bar{c}s$\\
		\bottomrule[1.0pt]
		\bottomrule[1.0pt]
	\end{tabular}
\end{table*}

\renewcommand\tabcolsep{0.65cm}
\renewcommand{\arraystretch}{1.60}
\begin{table*}[!htbp]
	\caption{The flaver wave functions of the octet hidden-charm molecular pentaquark states.}
	\label{TableFW1}
	\begin{tabular}{c|c|c}
		\toprule[1.0pt]
		\toprule[1.0pt]
		States&Flavors&Flavor wave functions\\
		\hline
		\multirow{2}{*}{$P_{\psi}^{N^+}$}&$8_{1f}$&$\frac{1}{\sqrt{3}}\Sigma_c^+\bar{D}^{(*)0}-\sqrt{\frac{2}{3}}\Sigma_c^{++}D^{(*)-}$\\
		&$8_{2f}$&$\Lambda_c^+\bar{D}^{(*)0}$\\
		\hline
		\multirow{2}{*}{$P_{\psi}^{N^0}$}&$8_{1f}$&$\frac{1}{\sqrt{3}}\Sigma_c^+D^{(*)-}-\sqrt{\frac{2}{3}}\Sigma_c^0\bar{D}^{(*)0}$\\
		&$8_{2f}$&$\Lambda_c^+D^{(*)-}$\\
		\hline
		\multirow{2}{*}{$P_{\psi_s}^{\Sigma^+}$}&$8_{1f}$&$\frac{1}{\sqrt{3}}\Xi_c^{\prime+}\bar{D}^{(*)0}-\sqrt{\frac{2}{3}}\Sigma_c^{++}D_s^{(*)-}$\\
		&$8_{2f}$&$\Xi_c^+\bar{D}^{(*)0}$\\
		\hline
		\multirow{2}{*}{$P_{\psi_s}^{\Sigma^0}$}&$8_{1f}$&$\frac{1}{\sqrt{6}}\Xi_c^{\prime+}D^{(*)-}+\frac{1}{\sqrt{6}}\Xi_c^{\prime0}\bar{D}^{(*)0}-\sqrt{\frac{2}{3}}\Sigma_c^+D_s^{(*)-}$\\
		&$8_{2f}$&$\frac{1}{\sqrt{2}}\Xi_c^+D^{(*)-}+\frac{1}{\sqrt{2}}\Xi_c^0\bar{D}^{(*)0}$\\
		\hline
		\multirow{2}{*}{$P_{\psi_s}^{\Lambda^0}$}&$8_{1f}$&$\frac{1}{\sqrt{2}}\Xi_c^{\prime+}D^{(*)-}-\frac{1}{\sqrt{2}}\Xi_c^{\prime0}\bar{D}^{(*)0}$\\
		&$8_{2f}$&$\frac{1}{\sqrt{6}}\Xi_c^+D^{(*)-}+\frac{1}{\sqrt{6}}\Xi_c^0\bar{D}^{(*)0}-\sqrt{\frac{2}{3}}\Lambda_c^+D_s^{(*)-}$\\
		\hline
		\multirow{2}{*}{$P_{\psi_s}^{\Sigma^-}$}&$8_{1f}$&$\frac{1}{\sqrt{3}}\Xi_c^{\prime0}D^{(*)-}-\sqrt{\frac{2}{3}}\Sigma_c^0D_s^{(*)-}$\\
		&$8_{2f}$&$\Xi_c^0D^{(*)-}$\\
		\hline
		\multirow{2}{*}{$P_{\psi_{ss}}^{N^0}$}&$8_{1f}$&$\frac{1}{\sqrt{3}}\Xi_c^{\prime+}D_s^{(*)-}-\sqrt{\frac{2}{3}}\Omega_c^0\bar{D}^{(*)0}$\\
		&$8_{2f}$&$\Xi_c^+D_s^{(*)-}$\\
		\hline
		\multirow{2}{*}{$P_{\psi_{ss}}^{N^-}$}&$8_{1f}$&$\frac{1}{\sqrt{3}}\Xi_c^{\prime0}D_s^{(*)-}-\sqrt{\frac{2}{3}}\Omega_c^0D^{(*)-}$\\
		&$8_{2f}$&$\Xi_c^0D_s^{(*)-}$\\
		\bottomrule[1.0pt]
		\bottomrule[1.0pt]
	\end{tabular}
\end{table*}

We construct the spin wave functions of the pentaquark states, the total angular momentum of the pentaquark states can be expressed as $J^P=J_b^P\otimes J_m^P$, where $J_b^P$ and $J_m^P$ are the spins of baryons and mesons, respectively. We consider the spin configurations $\frac{1}{2}^-(\frac{1}{2}^+\otimes0^-)$, $\frac{1}{2}^-(\frac{1}{2}^+\otimes1^-)$ and $\frac{3}{2}^-(\frac{1}{2}^+\otimes1^-)$ , the spin wave functions of the pentaquark states are listed in Table \ref{TableSW1}. Since $\phi_{\rm flavor}\chi_{\rm spin}=\rm symmetric$, the spin wave functions of the pentaquark consists of the spin wave functions of the charmed baryons and the anti-charmed mesons, we need to construct the symmetric and anti-symmetric spin wave functions of the constituent hadrons, the corresponding spin wave functions with different flavor representations are collected in Table \ref{TableSW2}. Here, $6_{\rm baryon}$, $\bar{3}_{\rm baryon}$ and $3_{\rm meson}$ represent the spin wave functions of the hadrons with different flavors.

\renewcommand\tabcolsep{0.65cm}
\renewcommand{\arraystretch}{1.60}
\begin{table*}[!htbp]
	\caption{The spin wave functions of the octet hidden-charm molecular pentaquark states.}
	\label{TableSW1}
	\begin{tabular}{c|c|c}
		\toprule[1.0pt]
		\toprule[1.0pt]
		States&$|S,S_3\rangle$&Spin wave functions\\
		\hline
		\multirow{3}{*}{$P_{\psi}$}&\multirow{2}{*}{$|\frac{1}{2},\frac{1}{2}\rangle$}&$|\frac{1}{2},\frac{1}{2}\rangle|0,0\rangle$\\
		&&$\sqrt{\frac{1}{3}}|\frac{1}{2},\frac{1}{2}\rangle|1,0\rangle-\sqrt{\frac{2}{3}}|\frac{1}{2},-\frac{1}{2}\rangle|1,1\rangle$\\
		\cline{2-3}
		&$|\frac{3}{2},\frac{1}{2}\rangle$&$\sqrt{\frac{2}{3}}|\frac{1}{2},\frac{1}{2}\rangle|1,0\rangle+\sqrt{\frac{1}{3}}|\frac{1}{2},-\frac{1}{2}\rangle|1,1\rangle$\\
		\bottomrule[1.0pt]
		\bottomrule[1.0pt]
	\end{tabular}
\end{table*}

\renewcommand\tabcolsep{0.65cm}
\renewcommand{\arraystretch}{1.60}
\begin{table*}[!htbp]
	\caption{The spin wave functions of constituent hadrons.}
	\label{TableSW2}
	\begin{tabular}{c|c|c}
		\toprule[1.0pt]
		\toprule[1.0pt]
		&$|S,S_3\rangle$&Spin wave functions\\
		\hline
		\multirow{2}{*}{$6_{\rm baryon}$}&$|\frac{1}{2},\frac{1}{2}\rangle$&$\sqrt{\frac{1}{6}}(2\uparrow\uparrow\downarrow-\downarrow\uparrow\uparrow-\uparrow\downarrow\uparrow)$\\
		&$|\frac{1}{2},-\frac{1}{2}\rangle$&$\sqrt{\frac{1}{6}}(\uparrow\downarrow\downarrow+\downarrow\uparrow\downarrow-2\downarrow\downarrow\uparrow)$\\
		\hline
		\multirow{2}{*}{$\bar{3}_{\rm baryon}$}&$|\frac{1}{2},\frac{1}{2}\rangle$&$\sqrt{\frac{1}{2}}(\uparrow\downarrow\uparrow-\downarrow\uparrow\uparrow)$\\
		&$|\frac{1}{2},-\frac{1}{2}\rangle$&$\sqrt{\frac{1}{2}}(\uparrow\downarrow\downarrow-\downarrow\uparrow\downarrow)$\\
		\hline
		\multirow{4}{*}{$3_{\rm meson}$}&$|0,0\rangle$&$\sqrt{\frac{1}{2}}(\uparrow\downarrow-\downarrow\uparrow)$\\
		&$|1,1\rangle$&$\uparrow\uparrow$\\
		&$|1,0\rangle$&$\sqrt{\frac{1}{2}}(\uparrow\downarrow+\downarrow\uparrow)$\\
		&$|1,-1\rangle$&$\downarrow\downarrow$\\
		\bottomrule[1.0pt]
		\bottomrule[1.0pt]
	\end{tabular}
\end{table*}

\section{Transition magnetic moments and radiative decay widths}\label{sec3}
In this section, we present the calculations of the transition magnetic moments of pentaquark states, and we calculate the radiative decay widths of pentaquark states through the results of transition magnetic moments. In the constituent quark model, the transition magnetic moments can be expressed as
\begin{eqnarray}
	\mu_{H\to {H}'}=\left\langle \psi_{H'}\right|\hat{\mu}_ze^{-i\bm{k}\cdot\bm{r}}|\psi_H\rangle,\label{tr}
\end{eqnarray}
where $\bm{k}$ is the emitted photon momentum, and $\psi_H$ and $\psi_{H'}$ denote the spin-flavor wave functions of the initial and final states, respectively. For the $S$-wave pentaquark state, the magnetic moments operator at the quark level can be expressed as
\begin{eqnarray}
	\hat{\mu}&=&\sum_{i}\frac{e_i}{2m_i}\hat{\sigma}_{i},\label{em}
\end{eqnarray}
where $e_i$ and $\sigma_i$ represent the charge and Pauli’s spin operator of the i-th constituent of the hadrons, respectively. When the emitted photon momentum is extremely small, the factor $\left\langle R_{i'}\right|e^{-i\bm{k}\cdot\bm{r}}|R_{i}\rangle$ is close to zero, and the spatial wave functions has almost no effect on the transition magnetic moments. The Eq. (\ref{tr}) can be simplified as
\begin{eqnarray}
	\mu_{H\to {H}'}=\left\langle \psi_{H'}\right|\hat{\mu}_z|\psi_H\rangle. \label{tmm1}
\end{eqnarray}

In order to calculate the transition magnetic moments of the $S$-wave pentaquark states, it is necessary to calculate the magnetic moments of the constituent hadrons and the transition magnetic moments between them. The calculations of the magnetic moments are similar to that of transition magnetic moments. Inserting the $z$ component of the magnetic moment operator into the corresponding spin-flavor wave functions, we obtain the magnetic moments
\begin{eqnarray}
	\mu_{{H}}&=&\left\langle\psi_H|\hat{\mu}_z\right|\psi_H\rangle, \label{mm}
\end{eqnarray}
we take the charmed baryon $\Sigma_c^+$ as an example to illustrate the calculations of the magnetic moments, the spin-flavor wave function of $\Sigma_c^+$ can be expressed as
\begin{eqnarray}
	\chi_{\Sigma_c^+}^{|\frac{1}{2},\frac{1}{2}\rangle}=\sqrt{\frac{1}{2}}(udc+duc)\otimes\sqrt{\frac{1}{6}}(2\uparrow\uparrow\downarrow-\downarrow\uparrow\uparrow-\uparrow\downarrow\uparrow).
\end{eqnarray}
According to Eq. (\ref{mm}), we can obtain the expression for the magnetic moment of $\Sigma_c^+$ state as $\frac{2}{3}\mu_u+\frac{2}{3}\mu_d-\frac{1}{3}\mu_c$. With the same method, we list the magnetic moments of the constituent hadrons in Table \ref{Tablemm1}. In the numerical calculations, we take the quark masses $m_u=m_d=0.336~{\rm GeV}$, $m_s=0.540~{\rm GeV}$, $m_c=1.660~{\rm GeV}$\cite{Wang:2018gpl}.

\renewcommand\tabcolsep{0.65cm}
\renewcommand{\arraystretch}{1.60}
\begin{table*}[!htbp]
	\caption{The magnetic moments of baryons and mesons, the unit is nuclear magnetic moment $\mu_N$.}
	\label{Tablemm1}
	\begin{tabular}{c|c|c|c}
		\toprule[1.0pt]
		\toprule[1.0pt]
		States&Quantities&Expressions&Results\\
		\hline
		\multirow{6}{*}{$6_f$}&$\Sigma_c^{++}$&$\frac{4}{3}\mu_u-\frac{1}{3}\mu_c$&2.36\\
		&$\Sigma_c^+$&$\frac{2}{3}\mu_u+\frac{2}{3}\mu_d-\frac{1}{3}\mu_c$&0.49\\
		&$\Sigma_c^0$&$\frac{4}{3}\mu_d-\frac{1}{3}\mu_c$&-1.37\\
		&$\Xi_c^{\prime+}$&$\frac{2}{3}\mu_u+\frac{2}{3}\mu_s-\frac{1}{3}\mu_c$&0.73\\
		&$\Xi_c^{\prime0}$&$\frac{2}{3}\mu_d+\frac{2}{3}\mu_s-\frac{1}{3}\mu_c$&-1.13\\
		&$\Omega_c^0$&$\frac{4}{3}\mu_s-\frac{1}{3}\mu_c$&-0.90\\
		\hline
		\multirow{3}{*}{$\bar{3}_f$}&$\Xi_c^+$&$\mu_c$&0.38\\
		&$\Xi_c^0$&$\mu_c$&0.38\\
		&$\Lambda_c^+$&$\mu_c$&0.38\\
		\hline
		\multirow{3}{*}{$3_f$}&$\bar{D}^{(*)0}$&$\mu_u+\mu_{\bar{c}}$&1.48\\
		&$D^{(*)-}$&$\mu_d+\mu_{\bar{c}}$&-1.31\\
		&$D_s^{(*)-}$&$\mu_s+\mu_{\bar{c}}$&-0.96\\
		\bottomrule[1.0pt]
		\bottomrule[1.0pt]
	\end{tabular}
\end{table*}

In the discussion that follows, we use $P_j(j=1,2,3)$ to represent octet hidden-charm pentaquark states with different spin configurations, and $P_1$, $P_2$ and $P_3$ represent pentaquark states with spin configurations $J^P=|\frac{1}{2}^-\rangle(\frac{1}{2}^+\otimes0^-)$, $J^P=|\frac{1}{2}^-\rangle(\frac{1}{2}^+\otimes1^-)$ and $J^P=|\frac{3}{2}^-\rangle(\frac{1}{2}^+\otimes1^-)$ respectively. We take the process $P^{N^+}_{2\psi}\to P^{N^+}_{1\psi}\gamma$ with $8_{1f}$ flavor representation as an example to illustrate the calculations of the transition magnetic moments of the $S$-wave hidden-charm pentaquark states. According to Tables \ref{TableFW1} and \ref{TableSW1}, we can write the spin-flavor wave functions of initial and final states
\begin{eqnarray}
	\chi_{P_2^{N^+}}&=&\left[\frac{1}{\sqrt{3}}\Sigma_c^+\bar{D}^{*0}-\sqrt{\frac{2}{3}}\Sigma_c^{++}D^{*-}\right]\\\nonumber
	&&\otimes\left[\sqrt{\frac{1}{3}}|\frac{1}{2},\frac{1}{2}\rangle|1,0\rangle-\sqrt{\frac{2}{3}}|\frac{1}{2},-\frac{1}{2}\rangle|1,1\rangle\right],\\
	\chi_{P_1^{N^+}}&=&\left[\frac{1}{\sqrt{3}}\Sigma_c^+\bar{D}^0-\sqrt{\frac{2}{3}}\Sigma_c^{++}D^-\right]\otimes|\frac{1}{2},\frac{1}{2}\rangle|0,0\rangle.
\end{eqnarray}
Taking the magnetic moment operator into the spin-flavor wave functions of initial and final states, we can calculate the transition magnetic moment of the process $P^{N^+}_{2\psi}\to P^{N^+}_{1\psi}\gamma$ is $\frac{\sqrt{3}}{9}(\mu_{\bar{D}^{*0}\to\bar{D}^0}+2\mu_{D^{*-}\to D^-}) $. We notice that the transition magnetic moments of pentaquark states are composed of the transition magnetic moments of its constituent hadrons, therefore we present the transition magnetic moment of the process $\bar{D}^{*0}\to\bar{D}^0\gamma$ as an example to illustrate the calculations, the spin-flavor wave functions of $\bar{D}^{*0}$ and $\bar{D}^0$ can be expressed as
\begin{eqnarray}
	\chi_{\bar{D}^{*0}}^{|1,0\rangle}&=\frac{1}{\sqrt{2}}|\bar{c}u \rangle\otimes|\uparrow \downarrow+\downarrow\uparrow\rangle,\\
	\chi_{\bar{D}^0}^{|0,0\rangle}&=\frac{1}{\sqrt{2}}|\bar{c}u \rangle\otimes|\uparrow \downarrow-\downarrow\uparrow\rangle.
\end{eqnarray}
Taking the above wave functions into Eq. (\ref{tmm1}), we calculate the transition magnetic moments of the $\bar{D}^{*0}\to\bar{D}^0\gamma$ process as $\mu_{\bar{c}}-\mu_u$, we can obtain the transition magnetic moments of the other constituent hadrons in Table \ref{TableTmm2}.

\renewcommand\tabcolsep{0.65cm}
\renewcommand{\arraystretch}{1.60}
\begin{table*}[!htbp]
	\caption{The transition magnetic moments of mesons, the unit is nuclear magnetic moment $\mu_N$.}
	\label{TableTmm2}
	\begin{tabular}{c|c|c}
		\toprule[1.0pt]
		\toprule[1.0pt]
		States&Expressions&Results\\
		\hline
		$\bar{D}^{*0}\to\bar{D}^0\gamma$&$\mu_{\bar{c}}-\mu_u$&-2.24\\
		$D^{*-}\to D^-\gamma$&$\mu_{\bar{c}}-\mu_d$&0.55\\
		$D_s^{*-}\to D_s^-\gamma$&$\mu_{\bar{c}}-\mu_s$&0.20\\
		\bottomrule[1.0pt]
		\bottomrule[1.0pt]
	\end{tabular}
\end{table*}

The linear combination of magnetic moments and transition magnetic moments enables us to calculate the transition magnetic moments of pentaquark states with the $8_{1f}$ and $8_{2f}$ flavor representations, we collect their results in Tables \ref{TableT1f} and \ref{TableT2f}.

Analyzing the above results for the transition magnetic moments of the hidden-charm pentaquark states, we can summarize the following key points :
\begin{itemize}
	\item The transition magnetic moments of the hidden-charm pentaquark states are not only related to the transition magnetic moments between their constituent hadrons, but also related to the magnetic moments of the hadrons.
	\item With different flavor representations, the decay processes have different transition magnetic moments. For example, for $P_{\psi}^{N^+}$ state, the transition magnetic moment of the process $\Sigma_cD^*|\frac{3}{2}^-\rangle\to \Sigma_cD|\frac{1}{2}^-\rangle\gamma$ with $8_{1f}$ flavor representation is -0.31, while the transition magnetic moment of the process $\Lambda_cD^*|\frac{3}{2}^-\rangle\to \Lambda_cD|\frac{1}{2}^-\rangle\gamma$ with $8_{2f}$ flavor representation is -1.83, since the pentaquark states have different flavor wave functions in different flavor representations.
	\item With the same flavor representations, the different transition magnetic moments satisfy proportional relations. For example, for state $P_{\psi}^{N^+}$ with $8_{1f}$ flavor representation, the transition magnetic moments satisfy the relation
	\begin{eqnarray}
	 \frac{\mu_{\Sigma_cD^*|\frac{3}{2}^-\rangle\to \Sigma_cD|\frac{1}{2}^-\rangle\gamma}}{\mu_{\Sigma_cD^*|\frac{1}{2}^-\rangle\to \Sigma_cD|\frac{1}{2}^-\rangle\gamma}}=\sqrt{2}.
	 \end{eqnarray}
\end{itemize}

\renewcommand\tabcolsep{0.00cm}
\renewcommand{\arraystretch}{1.80}
\begin{table*}[!htbp]
	\caption{The transition magnetic moments of pentaquark states with the $8_{1f}$ flavor representation. Here, the unit of the transition magnetic moment is nuclear magnetic moment $\mu_N$.}
	\label{TableT1f}
		\begin{tabular}{c|c|c|c}
			\toprule[1.0pt]
			\toprule[1.0pt]
			States&Processes&Expressions&Results\\
			\hline
			\multirow{3}{*}{$P_{\psi}^{N^+}$}&$\Sigma_cD^*|\frac{1}{2}^-\rangle\to \Sigma_cD|\frac{1}{2}^-\rangle\gamma$&$\frac{\sqrt{3}}{9}(\mu_{\bar{D}^{*0}\to\bar{D}^0}+2\mu_{D^{*-}\to D^-})$&-0.22\\
			&$\Sigma_cD^*|\frac{3}{2}^-\rangle\to \Sigma_cD|\frac{1}{2}^-\rangle\gamma$&$\frac{\sqrt{6}}{9}(\mu_{\bar{D}^{*0}\to\bar{D}^0}+2\mu_{D^{*-}\to D^-})$&-0.31\\
			&$\Sigma_cD^*|\frac{3}{2}^-\rangle\to \Sigma_cD^*|\frac{1}{2}^-\rangle\gamma$&$\frac{\sqrt{2}}{9}(2\mu_{\Sigma_c^+}+4\mu_{\Sigma_c^{++}})-\frac{\sqrt{2}}{9}(\mu_{\bar{D}^{*0}}+2\mu_{D^{*-}})$&1.81\\
			\hline
			\multirow{3}{*}{$P_{\psi}^{N^0}$}&$\Sigma_cD^*|\frac{1}{2}^-\rangle\to \Sigma_cD|\frac{1}{2}^-\rangle\gamma$&$\frac{\sqrt{3}}{9}(\mu_{D^{*-}\to D^-}+2\mu_{\bar{D}^{*0}\to\bar{D}^0})$&-0.76\\
			&$\Sigma_cD^*|\frac{3}{2}^-\rangle\to \Sigma_cD|\frac{1}{2}^-\rangle\gamma$&$\frac{\sqrt{6}}{9}(\mu_{D^{*-}\to D^-}+2\mu_{\bar{D}^{*0}\to\bar{D}^0})$&-1.07\\
			&$\Sigma_cD^*|\frac{3}{2}^-\rangle\to \Sigma_cD^*|\frac{1}{2}^-\rangle\gamma$&$\frac{\sqrt{2}}{9}(2\mu_{\Sigma_c^+}+4\mu_{\Sigma_c^0})-\frac{\sqrt{2}}{9}(\mu_{D^{*-}}+2\mu_{\bar{D}^{*0}})$&-0.97\\
			\hline
			\multirow{3}{*}{$P_{\psi_s}^{\Sigma^+}$}&$\Sigma_cD^*|\frac{1}{2}^-\rangle\to \Sigma_cD|\frac{1}{2}^-\rangle\gamma$&$\frac{\sqrt{3}}{9}(\mu_{\bar{D}^{*0}\to\bar{D}^0}+2\mu_{D_s^{*-}\to D_s^-})$&-0.35\\
			&$\Sigma_cD^*|\frac{3}{2}^-\rangle\to \Sigma_cD|\frac{1}{2}^-\rangle\gamma$&$\frac{\sqrt{6}}{9}(\mu_{\bar{D}^{*0}\to\bar{D}^0}+2\mu_{D_s^{*-}\to D_s^-})$&-0.50\\
			&$\Sigma_cD^*|\frac{3}{2}^-\rangle\to \Sigma_cD^*|\frac{1}{2}^-\rangle\gamma$&$\frac{\sqrt{2}}{9}(2\mu_{\Xi_c^{\prime+}}+4\mu_{\Sigma_c^{++}})-\frac{\sqrt{2}}{9}(\mu_{\bar{D}^{*0}}+2\mu_{D_s^{*-}})$&1.78\\
			\hline
			\multirow{3}{*}{$P_{\psi_s}^{\Sigma^0}$}&$\Sigma_cD^*|\frac{1}{2}^-\rangle\to \Sigma_cD|\frac{1}{2}^-\rangle\gamma$&$\frac{\sqrt{3}}{18}(\mu_{D^{*-}\to D^-}+\mu_{\bar{D}^{*0}\to\bar{D}^0})+\frac{2\sqrt{3}}{9}\mu_{D_s^{*-}\to D_s^-}$&-0.08\\
			&$\Sigma_cD^*|\frac{3}{2}^-\rangle\to \Sigma_cD|\frac{1}{2}^-\rangle\gamma$&$\frac{\sqrt{6}}{18}(\mu_{D^{*-}\to D^-}+\mu_{\bar{D}^{*0}\to\bar{D}^0})+\frac{2\sqrt{6}}{9}\mu_{D_s^{*-}\to D_s^-}$&-0.12\\
			&$\Sigma_cD^*|\frac{3}{2}^-\rangle\to \Sigma_cD^*|\frac{1}{2}^-\rangle\gamma$&$\frac{\sqrt{2}}{9}(\mu_{\Xi_c^{\prime+}}+\mu_{\Xi_c^{\prime0}}+4\mu_{\Sigma_c^+})-\frac{\sqrt{2}}{18}(\mu_{D^{*-}}+\mu_{\bar{D}^{*0}}+4\mu_{D_s^{*-}})$&0.53\\
			\hline
			\multirow{3}{*}{$P_{\psi_s}^{\Lambda^0}$}&$\Xi^{\prime}_cD^*|\frac{1}{2}^-\rangle\to \Xi^{\prime}_cD|\frac{1}{2}^-\rangle\gamma$&$\frac{\sqrt{3}}{6}(\mu_{D^{*-}\to D^-}+\mu_{\bar{D}^{*0}\to\bar{D}^0})$&-0.49\\
			&$\Xi^{\prime}_cD^*|\frac{3}{2}^-\rangle\to \Xi^{\prime}_cD|\frac{1}{2}^-\rangle\gamma$&$\frac{\sqrt{6}}{6}(\mu_{D^{*-}\to D^-}+\mu_{\bar{D}^{*0}\to\bar{D}^0})$&-0.69\\
			&$\Xi^{\prime}_cD^*|\frac{3}{2}^-\rangle\to \Xi^{\prime}_cD^*|\frac{1}{2}^-\rangle\gamma$&$\frac{\sqrt{2}}{3}(\mu_{\Xi_c^{\prime+}}+\mu_{\Xi_c^{\prime0}})-\frac{\sqrt{2}}{6}(\mu_{D^{*-}}+\mu_{\bar{D}^{*0}})$&-0.61\\
			\hline
			\multirow{3}{*}{$P_{\psi_s}^{\Sigma^-}$}&$\Sigma_cD^*|\frac{1}{2}^-\rangle\to \Sigma_cD|\frac{1}{2}^-\rangle\gamma$&$\frac{\sqrt{3}}{9}(\mu_{D^{*-}\to D^-}+2\mu_{D_s^{*-}\to D_s^-})$&0.18\\
			&$\Sigma_cD^*|\frac{3}{2}^-\rangle\to \Sigma_cD^*|\frac{1}{2}^-\rangle\gamma$&$\frac{\sqrt{6}}{9}(\mu_{D^{*-}\to D^-}+2\mu_{D_s^{*-}\to D_s^-})$&0.26\\
			&$\Sigma_cD^*|\frac{3}{2}^-\rangle\to \Sigma_cD^*|\frac{1}{2}^-\rangle\gamma$&$\frac{\sqrt{2}}{9}(2\mu_{\Xi_c^{\prime0}}+4\mu_{\Sigma_c^0})-\frac{\sqrt{2}}{9}(\mu_{D^{*-}}+2\mu_{D_s^{*-}})$&-0.71\\
			\hline
			\multirow{3}{*}{$P_{\psi_{ss}}^{N^0}$}&$\Omega_cD^*|\frac{1}{2}^-\rangle\to \Omega_cD|\frac{1}{2}^-\rangle\gamma$&$\frac{\sqrt{3}}{9}(\mu_{D_s^{*-}\to D_s^-}+2\mu_{\bar{D}^{*0}\to\bar{D}^0})$&-0.82\\
			&$\Omega_cD^*|\frac{3}{2}^-\rangle\to \Omega_cD|\frac{1}{2}^-\rangle\gamma$&$\frac{\sqrt{6}}{9}(\mu_{D_s^{*-}\to D_s^-}+2\mu_{\bar{D}^{*0}\to\bar{D}^0})$&-1.16\\
			&$\Omega_cD^*|\frac{3}{2}^-\rangle\to \Omega_cD^*|\frac{1}{2}^-\rangle\gamma$&$\frac{\sqrt{2}}{9}(2\mu_{\Xi_c^{\prime+}}+4\mu_{\Omega_c^0})-\frac{\sqrt{2}}{9}(\mu_{D_s^{*-}}+2\mu_{\bar{D}^{*0}})$&-0.65\\
			\hline
			\multirow{3}{*}{$P_{\psi_{ss}}^{N^-}$}&$\Omega_cD^*|\frac{1}{2}^-\rangle\to \Omega_cD|\frac{1}{2}^-\rangle\gamma$&$\frac{\sqrt{3}}{9}(\mu_{D_s^{*-}\to D_s^-}+2\mu_{D^{*-}\to D^-})$&0.25\\
			&$\Omega_cD^*|\frac{3}{2}^-\rangle\to \Omega^{\prime}_cD|\frac{1}{2}^-\rangle\gamma$&$\frac{\sqrt{6}}{9}(\mu_{D_s^{*-}\to D_s^-}+2\mu_{D^{*-}\to D^-})$&0.36\\
			&$\Omega_cD^*|\frac{3}{2}^-\rangle\to \Omega^{\prime}_cD^*|\frac{1}{2}^-\rangle\gamma$&$\frac{\sqrt{2}}{9}(2\mu_{\Xi_c^{\prime0}}+4\mu_{\Omega_c^0})-\frac{\sqrt{2}}{9}(\mu_{D_s^{*-}}+2\mu_{D^{*-}})$&-0.36\\
			\bottomrule[1.0pt]
			\bottomrule[1.0pt]
	\end{tabular}
\end{table*}

\renewcommand\tabcolsep{0.00cm}
\renewcommand{\arraystretch}{1.80}
\begin{table*}[!htbp]
	\caption{The transition magnetic moments of pentaquark states with the $8_{2f}$ flavor representation.}
	\label{TableT2f}
		\begin{tabular}{c|c|c|c}
			\toprule[1.0pt]
			\toprule[1.0pt]
			States&Processes&Expressions&Results\\
			\hline
			\multirow{3}{*}{$P_{\psi}^{N^+}$}&$\Lambda_cD^*|\frac{1}{2}^-\rangle\to \Lambda_cD|\frac{1}{2}^-\rangle\gamma$&$\frac{\sqrt{3}}{3}\mu_{\bar{D}^{*0}\to\bar{D}^0}$&-1.29\\
			&$\Lambda_cD^*|\frac{3}{2}^-\rangle\to \Lambda_cD|\frac{1}{2}^-\rangle\gamma$&$\frac{\sqrt{6}}{3}\mu_{\bar{D}^{*0}\to\bar{D}^0}$&-1.83\\
			&$\Lambda_cD^*|\frac{3}{2}^-\rangle\to \Lambda_cD^*|\frac{1}{2}^-\rangle\gamma$&$\frac{\sqrt{2}}{3}(2\mu_{\Lambda_c^+}-\mu_{\bar{D}^{*0}})$&-0.33\\
			\hline
			\multirow{3}{*}{$P_{\psi}^{N^0}$}&$\Lambda_cD^*|\frac{1}{2}^-\rangle\to \Lambda_cD|\frac{1}{2}^-\rangle\gamma$&$\frac{\sqrt{3}}{3}\mu_{D^{*-}\to D^-}$&0.32\\
			&$\Lambda_cD^*|\frac{3}{2}^-\rangle\to \Lambda_cD|\frac{1}{2}^-\rangle\gamma$&$\frac{\sqrt{6}}{3}\mu_{D^{*-}\to D^-}$&0.45\\
			&$\Lambda_cD^*|\frac{3}{2}^-\rangle\to \Lambda_cD^*|\frac{1}{2}^-\rangle\gamma$&$\frac{\sqrt{2}}{3}(2\mu_{\Lambda_c^+}-\mu_{D^{*-}})$&0.97\\
			\hline
			\multirow{3}{*}{$P_{\psi_s}^{\Sigma^+}$}&$\Xi_cD^*|\frac{1}{2}^-\rangle\to \Xi_cD|\frac{1}{2}^-\rangle\gamma$&$\frac{\sqrt{3}}{3}\mu_{\bar{D}^{*0}\to\bar{D}^0}$&-1.29\\
			&$\Xi_cD^*|\frac{3}{2}^-\rangle\to \Xi_cD|\frac{1}{2}^-\rangle\gamma$&$\frac{\sqrt{6}}{3}\mu_{\bar{D}^{*0}\to\bar{D}^0}$&-1.83\\
			&$\Xi_cD^*|\frac{3}{2}^-\rangle\to \Xi_cD^*|\frac{1}{2}^-\rangle\gamma$&$\frac{\sqrt{2}}{3}(2\mu_{\Xi_c^+}-\mu_{\bar{D}^{*0}})$&-0.34\\
			\hline
			\multirow{3}{*}{$P_{\psi_s}^{\Sigma^0}$}&$\Xi_cD^*|\frac{1}{2}^-\rangle\to \Xi_cD|\frac{1}{2}^-\rangle\gamma$&$\frac{\sqrt{3}}{6}(\mu_{D^{*-}\to D^-}+\mu_{\bar{D}^{*0}\to\bar{D}^0})$&-0.49\\
			&$\Xi_cD^*|\frac{3}{2}^-\rangle\to \Xi_cD|\frac{1}{2}^-\rangle\gamma$&$\frac{\sqrt{6}}{6}(\mu_{D^{*-}\to D^-}+\mu_{\bar{D}^{*0}\to\bar{D}^0})$&-0.69\\
			&$\Xi_cD^*|\frac{3}{2}^-\rangle\to \Xi_cD^*|\frac{1}{2}^-\rangle\gamma$&$\frac{\sqrt{2}}{3}(\mu_{\Xi_c^+}+\mu_{\Xi_c^0})-\frac{\sqrt{2}}{6}(\mu_{D^{*-}}+\mu_{\bar{D}^{*0}})$&0.31\\
			\hline
			\multirow{3}{*}{$P_{\psi_s}^{\Lambda^0}$}&$\Lambda_cD^*|\frac{1}{2}^-\rangle\to \Lambda_cD|\frac{1}{2}^-\rangle\gamma$&$\frac{\sqrt{3}}{18}(\mu_{D^{*-}\to D^-}+\mu_{\bar{D}^{*0}\to\bar{D}^0})+\frac{2\sqrt{3}}{9}\mu_{D_s^{*-}\to D_s^-}$&-0.08\\
			&$\Lambda_cD^*|\frac{3}{2}^-\rangle\to \Lambda_cD|\frac{1}{2}^-\rangle\gamma$&$\frac{\sqrt{6}}{18}(\mu_{D^{*-}\to D^-}+\mu_{\bar{D}^{*0}\to\bar{D}^0})+\frac{2\sqrt{6}}{9}\mu_{D_s^{*-}\to D_s^-}$&-0.12\\
			&$\Lambda_cD^*|\frac{3}{2}^-\rangle\to \Lambda_cD^*|\frac{1}{2}^-\rangle\gamma$&$\frac{\sqrt{2}}{9}(\mu_{\Xi_c^+}+\mu_{\Xi_c^0}+4\mu_{\Lambda_c^+})-\frac{\sqrt{2}}{18}(\mu_{D^{*-}}+\mu_{\bar{D}^{*0}}+4\mu_{D_s^{*-}})$&0.64\\
			\hline
			\multirow{3}{*}{$P_{\psi_s}^{\Sigma^-}$}&$\Xi_cD^*|\frac{1}{2}^-\rangle\to \Xi_cD|\frac{1}{2}^-\rangle\gamma$&$\frac{\sqrt{3}}{3}\mu_{D^{*-}\to D^-}$&0.32\\
			&$\Xi_cD^*|\frac{3}{2}^-\rangle\to \Xi_cD|\frac{1}{2}^-\rangle\gamma$&$\frac{\sqrt{6}}{3}\mu_{D^{*-}\to D^-}$&0.45\\
			&$\Xi_cD^*|\frac{3}{2}^-\rangle\to \Xi_cD^*|\frac{1}{2}^-\rangle\gamma$&$\frac{\sqrt{2}}{3}(2\mu_{\Xi_c^0}-\mu_{D^{*-}})$&0.97\\
			\hline
			\multirow{3}{*}{$P_{\psi_{ss}}^{N^0}$}&$\Xi_cD^*|\frac{1}{2}^-\rangle\to \Xi_cD|\frac{1}{2}^-\rangle\gamma$&$\frac{\sqrt{3}}{3}\mu_{D_s^{*-}\to D_s^-}$&0.12\\
			&$\Xi_cD^*|\frac{3}{2}^-\rangle\to \Xi_cD|\frac{1}{2}^-\rangle\gamma$&$\frac{\sqrt{6}}{3}\mu_{D_s^{*-}\to D_s^-}$&0.17\\
			&$\Xi_cD^*|\frac{3}{2}^-\rangle\to \Xi_cD^*|\frac{1}{2}^-\rangle\gamma$&$\frac{\sqrt{2}}{3}(2\mu_{\Xi_c^+}-\mu_{D_s^{*-}})$&0.81\\
			\hline
			\multirow{3}{*}{$P_{\psi_{ss}}^{N^-}$}&$\Xi_cD^*|\frac{1}{2}^-\rangle\to \Xi_cD|\frac{1}{2}^-\rangle\gamma$&$\frac{\sqrt{3}}{3}\mu_{D_s^{*-}\to D_s^-}$&0.12\\
			&$\Xi_cD^*|\frac{3}{2}^-\rangle\to \Xi_cD|\frac{1}{2}^-\rangle\gamma$&$\frac{\sqrt{6}}{3}\mu_{D_s^{*-}\to D_s^-}$&0.17\\
			&$\Xi_cD^*|\frac{3}{2}^-\rangle\to \Xi_cD^*|\frac{1}{2}^-\rangle\gamma$&$\frac{\sqrt{2}}{3}(2\mu_{\Xi_c^0}-\mu_{D_s^{*-}})$&0.81\\
			\bottomrule[1.0pt]
			\bottomrule[1.0pt]
	\end{tabular}
\end{table*}

With the transition magnetic moments we obtained in this work, we can calculate the radiative decay widths of the pentaquark states. The radiative decay widths and the transition magnetic moments satisfy relations, which can be expressed as\cite{Wang:2022nqs}
\begin{eqnarray}
	\Gamma_{H \to H^{\prime}\gamma}=\alpha_{\rm{EM}}\frac{E_{\gamma}^3}{3m_p^2} J_a\frac{\left|\mu_{H \to H^{\prime}}\right|^2}{\mu_N^2},\label{Wid}
\end{eqnarray}
where the electromagnetic fine structure constant $\alpha_{\rm{EM}}$ is $\frac{1}{137}$, $m_p$ is the the mass of proton with $m_p=0.938~{\rm GeV}$, the angular momentum coefficient $J_a(a=1,2,3)$ can be written
\begin{equation}
	\begin{cases}
	J_1=\frac{J_H+1}{J_H} &\text{when}~~~J_H=J_{H^{\prime}}\\
	J_2=J_H &\text{when}~~~J_H=J_{H^{\prime}}+1.\\
	J_3=\frac{(J_H+1)(2J_H+3)}{2J_H+1} &\text{when}~~~J_H=J_{H^{\prime}}-1\\
    \end{cases}
\end{equation}
For the $H\to H^{\prime}\gamma$ process, $E_{\gamma}$ is the momentum of the emitted photon, which can be written as 
\begin{eqnarray}
	E_{\gamma}=\frac{m^2_H-m^2_{H^\prime}}{2m_H},
\end{eqnarray}
where $m_H$ and $m_{H^\prime}$ represent the masses of the corresponding hadrons. Taking the results of the transition magnetic moments into the Eq. (\ref{Wid}), we can obtain the numerical results of the radiative decay widths in Tables \ref{TableWid}.

According to the results of the radiative decay widths in Table \ref{TableWid}, several noteworthy points can be identified: 
\begin{itemize}
\item For decay process $\Sigma_cD^*|\frac{3}{2}^-\rangle\to \Sigma_cD^*|\frac{1}{2}^-\rangle\gamma$, we can obtain its radiative decay width less than 0.06 {\rm keV}, which is due to the fact that the masses of $\Sigma_cD^*|\frac{3}{2}^-\rangle$ and $\Sigma_cD^*|\frac{1}{2}^-\rangle$ are very close to each other. In other flavor representations, the radiative decay widths of this group are also almost 0 due to the same reasons.
\item For decay processes $\Sigma_cD^*|\frac{1}{2}^-\rangle\to \Sigma_cD|\frac{1}{2}^-\rangle\gamma$ and $\Sigma_cD^*|\frac{3}{2}^-\rangle\to \Sigma_cD|\frac{1}{2}^-\rangle\gamma$, their transition magnetic moments are not close, but their radiative decay widths are quite close, which is due to the joint effects of the angular momentum coefficient $J_a$ and the transition magnetic moments in Eq. (\ref{Wid}). For example, for $P_{\psi}^{N^+}$ state with $8_{1f}$ flavor representation, their transition magnetic moments satisfy the relation
\begin{eqnarray}
	 \mu_{\Sigma_cD^*|\frac{3}{2}^-\rangle\to \Sigma_cD|\frac{1}{2}^-\rangle\gamma}=\sqrt{2}\mu_{\Sigma_cD^*|\frac{1}{2}^-\rangle\to \Sigma_cD|\frac{1}{2}^-\rangle\gamma},
\end{eqnarray}
while the coefficients of angular momentum $J_a$ satisfy the relation $\frac{J_2}{J_1}=\frac{1}{2}$. The relations lead to quite close radiative decay widths in Eq. (\ref{Wid}).
\item We notice that the radiative decay widths of the $P_{\psi}^{N^0}$ and $P_{\psi_{ss}}^{N^0}$ states in $8_{1f}$ flavor representation are larger than 12.0 {\rm keV}, the decay widths of the $P_{\psi}^{N^+}$ and $P_{\psi_s}^{\Sigma^+}$ states in $8_{2f}$ flavor representation are larger than 37.0 {\rm keV}, and the decay widths of the remaining states are around 5.0 {\rm keV}. This is helpful for distinguish the pentaquark states experimentally.
\end{itemize}

\renewcommand\tabcolsep{0.00cm}
\renewcommand{\arraystretch}{1.80}
\begin{table*}[!htbp]
	\caption{The radiative decay widths of pentaquark states with the $8_{1f}$ and $8_{2f}$ flavor representations. Here, the radiative decay width is in units of {\rm keV}.}
	\label{TableWid}
	\begin{tabular}{c|c|c|c|c}
		\toprule[1.0pt]
		\toprule[1.0pt]
		\multirow{2}{*}{States}&\multicolumn{2}{c}{$8_{1f}$}&\multicolumn{2}{|c}{$8_{2f}$}\\
		\cline{2-5}
		&Processes&Radiative decay widths&Processes&Radiative decay widths\\
		\hline
		\multirow{2}{*}{$P_{\psi}^{N^+}$}&$\Sigma_cD^*|\frac{1}{2}^-\rangle\to \Sigma_cD|\frac{1}{2}^-\rangle\gamma$&1.07&$\Lambda_cD^*|\frac{1}{2}^-\rangle\to \Lambda_cD|\frac{1}{2}^-\rangle\gamma$&37.61\\
		&$\Sigma_cD^*|\frac{3}{2}^-\rangle\to \Sigma_cD|\frac{1}{2}^-\rangle\gamma$&1.07&$\Lambda_cD^*|\frac{3}{2}^-\rangle\to \Lambda_cD|\frac{1}{2}^-\rangle\gamma$&37.84\\
		\hline
		\multirow{2}{*}{$P_{\psi}^{N^0}$}&$\Sigma_cD^*|\frac{1}{2}^-\rangle\to \Sigma_cD|\frac{1}{2}^-\rangle\gamma$&13.05&$\Lambda_cD^*|\frac{1}{2}^-\rangle\to \Lambda_cD|\frac{1}{2}^-\rangle\gamma$&2.25\\
		&$\Sigma_cD^*|\frac{3}{2}^-\rangle\to \Sigma_cD|\frac{1}{2}^-\rangle\gamma$&12.94&$\Lambda_cD^*|\frac{3}{2}^-\rangle\to \Lambda_cD|\frac{1}{2}^-\rangle\gamma$&2.22\\
		\hline
		\multirow{2}{*}{$P_{\psi_s}^{\Sigma^+}$}&$\Sigma_cD^*|\frac{1}{2}^-\rangle\to \Sigma_cD|\frac{1}{2}^-\rangle\gamma$&3.28&$\Xi_cD^*|\frac{1}{2}^-\rangle\to \Xi_cD|\frac{1}{2}^-\rangle\gamma$&37.69\\
		&$\Sigma_cD^*|\frac{3}{2}^-\rangle\to \Sigma_cD|\frac{1}{2}^-\rangle\gamma$&3.35&$\Xi_cD^*|\frac{3}{2}^-\rangle\to \Xi_cD|\frac{1}{2}^-\rangle\gamma$&37.92\\
		\hline
		\multirow{2}{*}{$P_{\psi_s}^{\Sigma^0}$}&$\Sigma_cD^*|\frac{1}{2}^-\rangle\to \Sigma_cD|\frac{1}{2}^-\rangle\gamma$&0.15&$\Xi_cD^*|\frac{1}{2}^-\rangle\to \Xi_cD|\frac{1}{2}^-\rangle\gamma$&5.36\\
		&$\Sigma_cD^*|\frac{3}{2}^-\rangle\to \Sigma_cD|\frac{1}{2}^-\rangle\gamma$&0.17&$\Xi_cD^*|\frac{3}{2}^-\rangle\to \Xi_cD|\frac{1}{2}^-\rangle\gamma$&5.31\\
		\hline
		\multirow{2}{*}{$P_{\psi_s}^{\Lambda^0}$}&$\Xi^{\prime}_cD^*|\frac{1}{2}^-\rangle\to \Xi^{\prime}_cD|\frac{1}{2}^-\rangle\gamma$&5.36&$\Lambda_cD^*|\frac{1}{2}^-\rangle\to \Lambda_cD|\frac{1}{2}^-\rangle\gamma$&0.15\\
		&$\Xi^{\prime}_cD^*|\frac{3}{2}^-\rangle\to \Xi^{\prime}_cD|\frac{1}{2}^-\rangle\gamma$&5.32&$\Lambda_cD^*|\frac{3}{2}^-\rangle\to \Lambda_cD|\frac{1}{2}^-\rangle\gamma$&0.17\\
		\hline
		\multirow{2}{*}{$P_{\psi_s}^{\Sigma^-}$}&$\Sigma_cD^*|\frac{1}{2}^-\rangle\to \Sigma_cD|\frac{1}{2}^-\rangle\gamma$&0.75&$\Xi_cD^*|\frac{1}{2}^-\rangle\to \Xi_cD|\frac{1}{2}^-\rangle\gamma$&2.25\\
		&$\Sigma_cD^*|\frac{3}{2}^-\rangle\to \Sigma_cD^*|\frac{1}{2}^-\rangle\gamma$&0.78&$\Xi_cD^*|\frac{3}{2}^-\rangle\to \Xi_cD|\frac{1}{2}^-\rangle\gamma$&2.23\\
		\hline
		\multirow{2}{*}{$P_{\psi_{ss}}^{N^0}$}&$\Omega_cD^*|\frac{1}{2}^-\rangle\to \Omega_cD|\frac{1}{2}^-\rangle\gamma$&15.46&$\Xi_cD^*|\frac{1}{2}^-\rangle\to \Xi_cD|\frac{1}{2}^-\rangle\gamma$&0.34\\
		&$\Omega_cD^*|\frac{3}{2}^-\rangle\to \Omega_cD|\frac{1}{2}^-\rangle\gamma$&15.47&$\Xi_cD^*|\frac{3}{2}^-\rangle\to \Xi_cD|\frac{1}{2}^-\rangle\gamma$&0.34\\
		\hline
		\multirow{2}{*}{$P_{\psi_{ss}}^{N^-}$}&$\Omega_cD^*|\frac{1}{2}^-\rangle\to \Omega_cD|\frac{1}{2}^-\rangle\gamma$&1.41&$\Xi_cD^*|\frac{1}{2}^-\rangle\to \Xi_cD|\frac{1}{2}^-\rangle\gamma$&0.34\\
		&$\Omega_cD^*|\frac{3}{2}^-\rangle\to \Omega^{\prime}_cD|\frac{1}{2}^-\rangle\gamma$&1.46&$\Xi_cD^*|\frac{3}{2}^-\rangle\to \Xi_cD|\frac{1}{2}^-\rangle\gamma$&0.34\\
		\bottomrule[1.0pt]
		\bottomrule[1.0pt]
	\end{tabular}
\end{table*}

\section{The axial-vector coupling constants of pentaquark states}\label{sec4}
In this section, we present the calculations of the axial-vector coupling constants of pentaquark states. The Lagrangian in the chiral quark model is
\begin{eqnarray}
	\mathscr{L}_{quark}&= \frac{1}{2}g_q\bar{\psi}_q\gamma^{\mu}\gamma_{5}\partial_{\mu}\phi\psi_q\sim\frac{1}{2}g_q\bar{\psi}_q\sigma_z\partial_z\phi\psi_q\nonumber\\
	&=\frac{1}{2}\frac{g_q}{f_{\pi}} (\bar{u}\sigma_z\partial_z\pi_0u-\bar{d}\sigma_z\partial_z\pi_0d) + ....,
\end{eqnarray}
where $g_q$ is the coupling coefficient, $f_{\pi}=92~{\rm MeV}$ is the decay constant of the meson, $\psi_q$ and $\bar{\psi}_q$ are the quark and antiquark fields, respectively. $\phi$ denotes the pseudoscalar meson field in the SU(2) flavor symmetry.
\begin{eqnarray}
	\phi = \frac{1}{f_{\pi}}
	\begin{pmatrix}
		\pi_0&\sqrt{2}\pi^+\\
		\sqrt{2}\pi^-&-\pi_0
	\end{pmatrix}.
\end{eqnarray}
The nucleon $N$ at quark level
\begin{eqnarray}		
	\left\langle N,j_3=+\frac{1}{2};\pi_0\right|\frac{1}{i}\frac{1}{2}\frac{g_q}{f_{\pi}}(\bar{u}\sigma_z\partial_z\pi_0u-\bar{d}\sigma_z\partial_z\pi_0d) \left|N,j_3=+\frac{1}{2}\right\rangle=\frac{5}{6}\frac{q_z}{f_{\pi}}g_q, \label{EQA3}
\end{eqnarray}
where $q_z$ is the external momentum of $\pi_0$, at the hadron level
\begin{eqnarray}		
	\left\langle N,j_3=+\frac{1}{2};\pi_0\right|\frac{1}{i}\frac{g_A}{f_{\pi}}\bar{N}\frac{{\textstyle\sum_{N_z}}}{2}\partial_{\mu}\phi N\left|N,j_3=+\frac{1}{2}\right\rangle=\frac{1}{2}\frac{q_z}{f_{\pi}}g_A, \label{EQA4}
\end{eqnarray}
where $g_A$ is the axial-vector charge of the nucleon. According to Eq. (\ref{EQA3}) and Eq. (\ref{EQA4}), we can obtain $g_q=\frac{3}{5}g_A$. 

The $SU(3)$ invariant Lagrangian of pentaquark state for $P_2(\frac{1}{2}^-)P_1(\frac{1}{2}^-)\pi_0$ in $8_{1f}$ flavor representation reads
\begin{align} 
	\mathscr{L}^{\frac{1}{2}^- \to \frac{1}{2}^-}_{\rho}
	&=Tr\Big(g_{\rho}\bar{P}_1 \gamma_{\mu}\gamma^{5}\{\partial^{\mu}\Phi,P_2\} +f_{\rho}\bar{P}_1\gamma_{\mu}\gamma^{5}[\partial^{\mu}\Phi,P_2]\Big),\label{equ:13}	
\end{align}
where $\{\partial^{\mu}\Phi,P_j\}=\partial^{\mu}\Phi P_j+P_j\partial^{\mu}\Phi$, $[\partial^{\mu}\Phi,P_j]=\partial^{\mu}\Phi P_j-P_j\partial^{\mu}\Phi$, and $g_{\rho}$, $f_{\rho}$ are independent axial-vector coupling constants of pentaquark states for $P_2(\frac{1}{2}^-)P_1(\frac{1}{2}^-)\pi_0$ in $8_{1f}$ flavor representation. Here, $P_j$ represent octet hidden-charm molecular pentaquark states with different spin configurations
\begin{equation} 
	\setlength{\arraycolsep}{0.1pt}
	P_{j}(j=1,2,3)= 
	\left(		
	\begin{array}{ccc}
		\frac{1}{\sqrt{2}}P_{\psi_s}^{\Sigma^0}+\frac{1}{\sqrt{6}}P_{\psi_s}^{\Lambda^0}
		&P_{\psi_s}^{\Sigma^+}
		&P_{\psi}^{N^+}
		\\
		P_{\psi_s}^{\Sigma^-}
		&-\frac{1}{\sqrt{2}}P_{\psi_s}^{\Sigma^0}+\frac{1}{\sqrt{6}}P_{\psi_s}^{\Lambda^0}
		&P_{\psi}^{N^0}
		\\
		P_{\psi_{ss}}^{N^-}
		&P_{\psi_{ss}}^{N^0}
		&\frac{2}{\sqrt{3}}P_{\psi_s}^{\Lambda^0}
		\\
	\end{array}
	\right).\label{equ:20}
\end{equation}
$\Phi$ represents the pseudoscalar meson field in $SU(3)$ flavor symmetry
\begin{align} 
	\Phi = \sum_{a=1}^{8}\lambda_{a}\Phi_{a} \equiv
	\left(
	\begin{array}{ccc}
		\pi_0 + \frac{1}{\sqrt{3}}\eta
		&\sqrt{2}\pi^+
		&\sqrt{2}K^+
		\\
		\sqrt{2}\pi^-
		&-\pi_0 + \frac{1}{\sqrt{3}}\eta
		&\sqrt{2}K^0
		\\	
		\sqrt{2}K^-
		&\sqrt{2}\bar{K}^0
		&-\frac{2}{\sqrt{3}}\eta
	\end{array}
	\right).
\end{align}
Take $\Phi$, $P_j$ into Eq. (\ref{equ:13}) and expand it. Here, we merely consider $\pi_0$ term, we obtain
\begin{align}
	\mathscr{L}_{\rho}^{\frac{1}{2}^-\to \frac{1}{2}^-}
	&={\frac{2}{\sqrt{3}}}{\frac{1}{f_{\pi}}} g_{\rho} (\bar{P}_1^{\Sigma^0}\Sigma_{z}\partial_{z}{\pi^0}P_2^{\Lambda^0}
	+\bar{P}_1^{\Lambda^0}\Sigma_{z}\partial_{z}{\pi^0}P_2^{\Sigma^0})\nonumber\\
	&-{\frac{1}{f_{\pi}}}(f_{\rho}+g_{\rho})\bar{P}_1^{N^0}\Sigma_{z}\partial_{z}{\pi^0}{P_2^{N^0}}\nonumber\\
	&+{\frac{1}{f_{\pi}}}(g_{\rho}+f_{\rho})\bar{P}_1^{N^+}\Sigma_{z}\partial_{z}{\pi^0}P_2^{N^+}\nonumber\\
	&+{\frac{1}{f_{\pi}}}(g_{\rho}-f_{\rho})\bar{P}_1^{N^-}\Sigma_{z}\partial_{z}{\pi^0}P_2^{N^-}\nonumber\\
	&+{\frac{1}{f_{\pi}}}(f_{\rho}-g_{\rho})\bar{P}_1^{N^0}\Sigma_z\partial_{z}{\pi^0}P_2^{N^0}\nonumber\\
	&+{\frac{1}{f_{\pi}}}2f_{\rho}\bar{P}_1^{\Sigma^+}\Sigma_z\partial_{z}{\pi^0}P_2^{\Sigma^+}\nonumber\\
	&-{\frac{1}{f_{\pi}}}2f_{\rho}\bar{P}_1^{\Sigma^-}\Sigma_z\partial_{z}{\pi^0}P_2^{\Sigma^-}.\label{equ:21}
\end{align}

At the hadron level, for $P_{\psi}^{N^+}$ state with $8_{1f}$ flavor representation, the Lagrangian for the decay process $P_2\to P_1\pi_0$ reads
\begin{align} 
	\mathscr{L}^{\frac{1}{2}^-\to \frac{1}{2}^-}_{P_{\psi}^{N^+}}
	&=
	\frac{1}{2}(g_{\rho}+f_{\rho})\bar{P}_1^{N^+}\gamma^{\mu}\gamma_5\partial_{\mu}\Phi P_2^{N^+}
	\nonumber  \\
	&\sim 
	\frac{g_{\rho}+f_{\rho}}{f_\pi}\bar{P}_1^{N^+}\frac{\Sigma_z}{2}\partial_{z}  {\pi_0}P_2^{N^+}. 
\end{align}
For $P_{\psi_{ss}}^{N^0}$ state with $8_{1f}$ flavor representation, the Lagrangian for the process $P_2\to P_1\pi_0$ reads
\begin{align} 
	\mathscr{L}^{\frac{1}{2}^-\to \frac{1}{2}^-}_{P_{\psi_{ss}}^{N^0}}
	&=
	\frac{1}{2}(f_{\rho}-g_{\rho})\bar{P}_1^{N^0}\gamma^{\mu}\gamma_5\partial_{\mu}\Phi P_2^{N^0} 
	\nonumber  \\
	&\sim 
	\frac{f_{\rho}-g_{\rho}}{f_\pi}\bar{P}_1^{N^0}\frac{\Sigma_z}{2}\partial_{z}{\pi_0}P_2^{N^0},
\end{align}
where $\frac{\Sigma_z}{2}$ is the spin operator of the hidden-charm pentaquark states. At the hadron level, the axial-vector coupling constants read
\begin{align} 	
	\left \langle P_{\psi}^{N^+}; \pi_0\right|\frac{g_{\rho}+f_{\rho}}{f_\pi}\bar{P}_{\psi}^{N^+}\frac{\Sigma_z}{2}\partial_{z}  {\pi_0}P_{\psi}^{N^+}\left|{P_{\psi}^{N^+}} 
	\right\rangle&=\frac{g_{\rho}+f_{\rho}}{2}\frac{q_z}{f_\pi},\\	
	\left\langle P_{\psi_{ss}}^{N^0}; \pi_0\right|\frac{f_{\rho}-g_{\rho}}{f_\pi}\bar{P}_{\psi_{ss}}^{N^0}\frac{\Sigma_z}{2}\partial_{z} {\pi_0}P_{\psi  ss}^{N^0}\left|P_{\psi_{ss}}^{N^0}
	\right\rangle&=\frac{f_{\rho}-g_{\rho}}{2}\frac{q_z}{f_\pi}.
\end{align}

At the quark level, we notice that the $\pi$ meson interactions only exist between light quarks. We take $P_{\psi}^{N^+}$ state with $8_{1f}$ flavor representation as example to illustrate the calculations, its wave function is
\begin{eqnarray}
	\left|P_{\psi}^{N^+}\right\rangle&=&\left|\sqrt{\frac{1}{6}}(udc)(\bar{c}u)+\sqrt{\frac{1}{6}}(duc)(\bar{c}u)-\sqrt{\frac{2}{3}}(uuc)(\bar{c}d)\right\rangle\\
	&&\otimes\left|\big[(q_1q_2)_{s_1}\otimes c\big]_{s_2}\otimes(\bar{c}q_3)_{s_3}\right\rangle^{J_z}_J,
\end{eqnarray}
where $s_1$ is the spin of the light diquark $q_1q_2$, which couples to the spin of charm quark to form the spin $s_2$. $s_3$ is the spin of the meson $\bar{c}q_3$, which couples to the baryon spin $s_2$ to form the total angular momentum $J$, and $J_z$ is its third component. For the process $P_2\to P_1\pi_0$, the wave functions of initial and final states can be written as
\begin{eqnarray}
	\left|P_2^{N^+}\right\rangle&=&\left|\sqrt{\frac{1}{6}}(udc)(\bar{c}u)+\sqrt{\frac{1}{6}}(duc)(\bar{c}u)-\sqrt{\frac{2}{3}}(uuc)(\bar{c}d)\right\rangle\\\nonumber
	&&\otimes\left[\frac{1}{6}(2|\uparrow\uparrow\downarrow\rangle-|\downarrow\uparrow\uparrow\rangle-|\uparrow\downarrow\uparrow\rangle)(|\uparrow\downarrow\rangle+|\downarrow\uparrow\rangle)-\frac{1}{3}(|\uparrow\downarrow\downarrow\rangle+|\downarrow\uparrow\downarrow\rangle-2|\downarrow\downarrow\uparrow\rangle)|\uparrow\uparrow\rangle\right],\\
	\left|P_1^{N^+}\right\rangle&=&\left|\sqrt{\frac{1}{6}}(udc)(\bar{c}u)+\sqrt{\frac{1}{6}}(duc)(\bar{c}u)-\sqrt{\frac{2}{3}}(uuc)(\bar{c}d)\right\rangle\\\nonumber
	&&\otimes\left[\frac{\sqrt{3}}{6}(2|\uparrow\uparrow\downarrow\rangle-|\downarrow\uparrow\uparrow\rangle-|\uparrow\downarrow\uparrow\rangle)(|\uparrow\downarrow\rangle-|\downarrow\uparrow\rangle)\right].
\end{eqnarray}
For the above $P_{\psi}^{N^+}$ molecular state, its axial-vector coupling constant at the quark level can be expressed as
\begin{eqnarray}
	&&\left\langle P_1^{N^+},+\frac{1}{2};\pi_0\right|\frac{1}{i}\frac{1}{2}\frac{g_q}{f_{\pi}}(\bar{u}\sigma_z\partial_z\pi_0u-\bar{d}\sigma_z\partial_z\pi_0d)\left|P_2^{N^+},+\frac{1}{2}\right\rangle\nonumber\\
	&=&\frac{1}{2}(-\frac{\sqrt{3}}{18}-\frac{\sqrt{3}}{18}+\frac{2\sqrt{3}}{9})\frac{q_z}{f_{\pi}}g_q=\frac{\sqrt{3}}{18}\frac{q_z}{f_{\pi}}g_q. \label{QL}
\end{eqnarray}
Additionally, the axial-vector coupling constant can also be expressed in another way, and Eq. (\ref{QL}) can be written as
\begin{eqnarray}
	&&\left\langle P_1^{N^+},+\frac{1}{2};\pi_0\right|\frac{1}{i}\frac{1}{2}\frac{g_q}{f_{\pi}}(\bar{u}\sigma_z\partial_z\pi_0u-\bar{d}\sigma_z\partial_z\pi_0d)\left|P_2^{N^+},+\frac{1}{2}\right\rangle\nonumber\\
	&=&\frac{\sqrt{3}}{9}\left\langle \bar{D}^0;\pi_0\right|\frac{1}{i}\frac{1}{2}\frac{g_q}{f_{\pi}}(\bar{u}\sigma_z\partial_z\pi_0u-\bar{d}\sigma_z\partial_z\pi_0d)\left|\bar{D}^{*0}\right\rangle\nonumber\\
	&&+\frac{2\sqrt{3}}{9}\left\langle \bar{D}^-;\pi_0\right|\frac{1}{i}\frac{1}{2}\frac{g_q}{f_{\pi}}(\bar{u}\sigma_z\partial_z\pi_0u-\bar{d}\sigma_z\partial_z\pi_0d)\left|\bar{D}^{*-}\right\rangle\nonumber\\
	&=&\frac{\sqrt{3}}{18}\frac{q_z}{f_{\pi}}g_q. 
\end{eqnarray}
With the same method, the axial-vector coupling constant for the process $P_2\to P_1\pi_0$ of ${P_{\psi_{ss}}^{N^0}}$ state read
\begin{align} 
	\left\langle P_1^{N^0} , +\frac{1}{2} ;~\pi_0\right| \mathscr{L}_{quark}\left|P_2^{N^0},
	+\frac{1}{2}\right\rangle
	=-\frac{\sqrt{3}}{9} \frac{q_z}{f_\pi}g_q.\label{equ:28}
\end{align}
Compare with the axial-vector charge of the nucleon
\begin{align} 
	\frac{\frac{1}{2}g_{A}}{\frac{5}{6}g_{q}}	 
	&= 
	\frac{\frac{g_{\rho}+f_{\rho}}{2}}{\frac{\sqrt{3}}{18}g_{q}} 
	=
	\frac{\frac{f_{\rho}-g_{\rho}}{2}}{-\frac{\sqrt{3}}{9}g_{q}}. \label{equ:28}
\end{align}
We obtain $g_{\rho}=\frac{\sqrt{3}}{10}g_{A}$ and $f_{\rho}=-\frac{\sqrt{3}}{30}g_{A}$. We can calculate the axial-vector coupling constants for the octet pentaquark states via the combination of the coupling constants $g_{\rho}$ and $f_{\rho}$. With the same methods, we obtain the axial-vector coupling constants of the pentaquark states with other flavor representations. 

The Lagrangian of pentaquark states for the process $P_3\to P_1\pi_0$ in $8_{1f}$ flavor representation reads
\begin{align} 
	\mathscr{L}^{\frac{3}{2}^- \to \frac{1}{2}^-}_{\alpha}
	&=Tr\Big(g_{\alpha}\bar{P}_1 \gamma_{\mu}\gamma^{5}\{\partial^{\mu}\Phi,P_{3\nu}\} +f_{\alpha}\bar{P}_1\gamma_{\mu}\gamma^{5}[\partial^{\mu}\Phi,P_{3\nu}]\Big).
\end{align}
The Lagrangian of pentaquark states for the process $P_3\to P_2\pi_0$ in $8_{1f}$ flavor representation reads
\begin{align} 
	\mathscr{L}^{\frac{3}{2}^- \to \frac{1}{2}^-}_{\beta}
	&=Tr\Big(g_{\beta}\bar{P}_2 \gamma_{\mu}\gamma^{5}\{\partial^{\mu}\Phi,P_{3\nu}\} +f_{\beta}\bar{P}_2\gamma_{\mu}\gamma^{5}[\partial^{\mu}\Phi,P_{3\nu}]\Big).
\end{align}
The Lagrangian of pentaquark states for the process $P_2\to P_1\pi_0$ in $8_{2f}$ flavor representation reads
\begin{align} 
	\mathscr{L}^{\frac{1}{2}^- \to \frac{1}{2}^-}_{\kappa}
	&=Tr\Big(g_{\kappa}\bar{P}_1 \gamma_{\mu}\gamma^{5}\{\partial^{\mu}\Phi,P_2\} +f_{\kappa}\bar{P}_1\gamma_{\mu}\gamma^{5}[\partial^{\mu}\Phi,P_2]\Big).
\end{align}
The Lagrangian of pentaquark states for the process $P_3\to P_1\pi_0$ in $8_{2f}$ flavor representation reads
\begin{align} 
	\mathscr{L}^{\frac{3}{2}^- \to \frac{1}{2}^-}_{\tau}
	&=Tr\Big(g_{\tau}\bar{P}_1 \gamma_{\mu}\gamma^{5}\{\partial^{\mu}\Phi,P_{3\nu}\} +f_{\tau}\bar{P}_1\gamma_{\mu}\gamma^{5}[\partial^{\mu}\Phi,P_{3\nu}]\Big).
\end{align}
The Lagrangian of pentaquark states for the process $P_3\to P_2\pi_0$ in $8_{2f}$ flavor representation reads
\begin{align} 
	\mathscr{L}^{\frac{3}{2}^- \to \frac{1}{2}^-}_{\omega}
	&=Tr\Big(g_{\omega}\bar{P}_2 \gamma_{\mu}\gamma^{5}\{\partial^{\mu}\Phi,P_{3\nu}\} +f_{\omega}\bar{P}_2\gamma_{\mu}\gamma^{5}[\partial^{\mu}\Phi,P_{3\nu}]\Big).
\end{align}
With the same method as above, we can calculate the values of the coupling constants $g_i$ and $f_i$ ($i=\rho,\alpha,\beta,\kappa,\tau,\omega$) of the pentaquark states with different flavor representations, and we list them in Table \ref{Tablegf}. With the coupling constants $g_i$ and $f_i$, we can obtain the axial-vector coupling constants of the octet hidden-charm pentaquark in the $8_{1f}$ and $8_{2f}$ flavor representations, and their results are listed in Table \ref{TableAx1} and Table \ref{TableAx2} respectively.
\renewcommand\tabcolsep{0.65cm}
\renewcommand{\arraystretch}{1.60}
\begin{table*}[!htbp]
	\caption{The transition coupling constants of the pentaquark states with different flavor representations.}
	\label{Tablegf}
	\begin{tabular}{c|c|c|c}
		\toprule[1.0pt]
		\toprule[1.0pt]
		Constants&Results&Constants&Results\\
		\hline
		$g_{\rho}$&$\frac{\sqrt{3}}{10}g_A$&$f_{\rho}$&$-\frac{\sqrt{3}}{30}g_A$\\
		$g_{\alpha}$&$\frac{3}{10}g_A$&$f_{\alpha}$&$-\frac{1}{10}g_A$\\
		$g_{\beta}$&$\frac{7\sqrt{3}}{30}g_A$&$f_{\beta}$&$\frac{17\sqrt{3}}{90}g_A$\\
		$g_{\kappa}$&$-\frac{\sqrt{3}}{10}g_A$&$f_{\kappa}$&$-\frac{\sqrt{3}}{10}g_A$\\
		$g_{\tau}$&$-\frac{3}{10}g_A$&$f_{\tau}$&$-\frac{3}{10}g_A$\\
		$g_{\omega}$&$-\frac{\sqrt{3}}{10}g_A$&$f_{\omega}$&$-\frac{\sqrt{3}}{10}g_A$\\
		\bottomrule[1.0pt]
		\bottomrule[1.0pt]
	\end{tabular}
\end{table*}

Analyzing the numerical results in Table \ref{TableAx1} and Table \ref{TableAx2}, we can summarize the several points :\\
({\romannumeral1})Comparing with our previous work\cite{Li:2024wxr}, we notice that the axial-vector coupling constants of the pentaquark states are generally smaller than that of the nucleon $g_A$, which is due to the fact that the contributions of the three light quarks to the axial-vector coupling constants interfere with each other, resulting in the smaller axial-vector coupling constants.\\
({\romannumeral2})The axial-vector coupling constants of the pentaquark states with different flavor representations are not the same in the identical decay modes. With different flavor representations, the pentaquark states with the same angular momentum $J$ have different spin-flavor wave functions, and the axial-vector coupling constants of the hidden-charm pentaquark states are sensitive to the flavor and spin configurations.\\
({\romannumeral3})For the $P_{\psi_{ss}}^{N^0}$, $P_{\psi_{ss}}^{N^-}$ states with the $8_{2f}$ flavor representation, due to the coupling constants $g_{\kappa}=f_{\kappa}$, $g_{\tau}=f_{\tau}$ and $g_{\omega}=f_{\omega}$, the axial-vector coupling constants of the processes $P_{\psi}|\frac{1}{2}^-\rangle\to P_{\psi}|\frac{1}{2}^-\rangle\pi_0$, $P_{\psi}|\frac{3}{2}^-\rangle\to P_{\psi}|\frac{1}{2}^-\rangle\pi_0$ are equal to 0.\\
\renewcommand\tabcolsep{0.00cm}
\renewcommand{\arraystretch}{1.80}
\begin{table*}[!htbp]
	\caption{The axial-vector coupling constants of the octet hidden-charm pentaquark states with the $8_{1f}$ flavor representation.}
	\label{TableAx1}
		\begin{tabular}{c|c|c|c|c}
			\toprule[1.0pt]
			\toprule[1.0pt]
			Processes&&&Coefficients&Results\\
			\hline
			\multirow{8}{*}{$\frac{1}{2}^-\to\frac{1}{2}^-$}&$P_{\psi}^{N^+}$&$\left\langle \Sigma_cD, ^{2}S_{\frac{1}{2}}\right|\mathscr{L}\left|\Sigma_cD^*, ^{2}S_{\frac{1}{2}}\right\rangle$&$g_{\rho}+f_{\rho}$&$\frac{\sqrt{3}}{15}g_A$\\
			&$P_{\psi}^{N^0}$&$\left\langle \Sigma_cD, ^{2}S_{\frac{1}{2}}\right|\mathscr{L}\left|\Sigma_cD^*, ^{2}S_{\frac{1}{2}}\right\rangle$&$-g_{\rho}-f_{\rho}$&$-\frac{\sqrt{3}}{15}g_A$\\
			&$P_{\psi_s}^{\Sigma^+}$&$\left\langle \Sigma_cD, ^{2}S_{\frac{1}{2}}\right|\mathscr{L}\left|\Sigma_cD^*, ^{2}S_{\frac{1}{2}}\right\rangle$&$2f_{\rho}$&$-\frac{\sqrt{3}}{15}g_A$\\
			&$P_{\psi_s}^{\Sigma^0}$&$\left\langle \Sigma_cD, ^{2}S_{\frac{1}{2}}\right|\mathscr{L}\left|\Sigma_cD^*, ^{2}S_{\frac{1}{2}}\right\rangle$&$\frac{2}{\sqrt{3}}g_{\rho}$&$\frac{1}{5}g_A$\\
			&$P_{\psi_s}^{\Lambda^0}$&$\left\langle \Xi^{\prime}_cD, ^{2}S_{\frac{1}{2}}\right|\mathscr{L}\left|\Xi^{\prime}_cD^*, ^{2}S_{\frac{1}{2}}\right\rangle$&$\frac{2}{\sqrt{3}}g_{\rho}$&$\frac{1}{5}g_A$\\
			&$P_{\psi_s}^{\Sigma^-}$&$\left\langle \Xi^{\prime}_cD, ^{2}S_{\frac{1}{2}}\right|\mathscr{L}\left|\Xi^{\prime}_cD^*, ^{2}S_{\frac{1}{2}}\right\rangle$&$-2f_{\rho}$&$-\frac{\sqrt{3}}{15}g_A$\\
			&$P_{\psi_{ss}}^{N^0}$&$\left\langle \Omega_cD, ^{2}S_{\frac{1}{2}}\right|\mathscr{L}\left|\Omega_cD^*, ^{2}S_{\frac{1}{2}}\right\rangle$&$-g_{\rho}+f_{\rho}$&$-\frac{2\sqrt{3}}{15}g_A$\\
			&$P_{\psi_{ss}}^{N^-}$&$\left\langle \Omega_cD, ^{2}S_{\frac{1}{2}}\right|\mathscr{L}\left|\Omega_cD^*, ^{2}S_{\frac{1}{2}}\right\rangle$&$g_{\rho}-f_{\rho}$&$\frac{2\sqrt{3}}{15}g_A$\\
			\hline
			Processes&&&Coefficients&Results\\
			\hline
			\multirow{16}{*}{$\frac{3}{2}^-\to\frac{1}{2}^-$}&\multirow{2}{*}{$P_{\psi}^{N^+}$}&$\left\langle \Sigma_cD, ^{2}S_{\frac{1}{2}}\right|\mathscr{L}\left|\Sigma_cD^*, ^{4}S_{\frac{3}{2}}\right\rangle$&$g_{\alpha}+f_{\alpha}$&$\frac{1}{5}g_A$\\
			&&$\left\langle \Sigma_cD^*, ^{2}S_{\frac{1}{2}}\right|\mathscr{L}\left|\Sigma_cD^*, ^{4}S_{\frac{3}{2}}\right\rangle$&$g_{\beta}+f_{\beta}$&$\frac{19\sqrt{3}}{45}g_A$\\
			&\multirow{2}{*}{$P_{\psi}^{N^0}$}&$\left\langle \Sigma_cD, ^{2}S_{\frac{1}{2}}\right|\mathscr{L}\left|\Sigma_cD^*, ^{4}S_{\frac{3}{2}}\right\rangle$&$-g_{\alpha}-f_{\alpha}$&$-\frac{1}{5}g_A$\\
			&&$\left\langle \Sigma_cD^*, ^{2}S_{\frac{1}{2}}\right|\mathscr{L}\left|\Sigma_cD^*, ^{4}S_{\frac{3}{2}}\right\rangle$&$-g_{\beta}-f_{\beta}$&$-\frac{19\sqrt{3}}{45}g_A$\\
			&\multirow{2}{*}{$P_{\psi_s}^{\Sigma^+}$}&$\left\langle \Sigma_cD, ^{2}S_{\frac{1}{2}}\right|\mathscr{L}\left|\Sigma_cD^*, ^{4}S_{\frac{3}{2}}\right\rangle$&$2f_{\alpha}$&$-\frac{1}{5}g_A$\\
			&&$\left\langle \Sigma_cD^*, ^{2}S_{\frac{1}{2}}\right|\mathscr{L}\left|\Sigma_cD^*, ^{4}S_{\frac{3}{2}}\right\rangle$&$2f_{\beta}$&$\frac{17\sqrt{3}}{45}g_A$\\
			&\multirow{2}{*}{$P_{\psi_s}^{\Sigma^0}$}&$\left\langle \Sigma_cD, ^{2}S_{\frac{1}{2}}\right|\mathscr{L}\left|\Sigma_cD^*, ^{4}S_{\frac{3}{2}}\right\rangle$&$\frac{2}{\sqrt{3}}g_{\alpha}$&$\frac{\sqrt{3}}{5}g_A$\\
			&&$\left\langle \Sigma_cD^*, ^{2}S_{\frac{1}{2}}\right|\mathscr{L}\left|\Sigma_cD^*, ^{4}S_{\frac{3}{2}}\right\rangle$&$\frac{2}{\sqrt{3}}g_{\beta}$&$\frac{21}{45}g_A$\\
			&\multirow{2}{*}{$P_{\psi_s}^{\Lambda^0}$}&$\left\langle \Xi^{\prime}_cD, ^{2}S_{\frac{1}{2}}\right|\mathscr{L}\left|\Xi^{\prime}_cD^*, ^{4}S_{\frac{3}{2}}\right\rangle$&$\frac{2}{\sqrt{3}}g_{\alpha}$&$\frac{\sqrt{3}}{5}g_A$\\
			&&$\left\langle \Xi^{\prime}_cD^*, ^{2}S_{\frac{1}{2}}\right|\mathscr{L}\left|\Xi^{\prime}_cD^*, ^{4}S_{\frac{3}{2}}\right\rangle$&$\frac{2}{\sqrt{3}}g_{\beta}$&$\frac{16}{45}g_A$\\
			&\multirow{2}{*}{$P_{\psi_s}^{\Sigma^-}$}&$\left\langle \Sigma_cD, ^{2}S_{\frac{1}{2}}\right|\mathscr{L}\left|\Sigma_cD^*, ^{4}S_{\frac{3}{2}}\right\rangle$&$-2f_{\alpha}$&$\frac{1}{5}g_A$\\
			&&$\left\langle \Sigma_cD^*, ^{2}S_{\frac{1}{2}}\right|\mathscr{L}\left|\Sigma_cD^*, ^{4}S_{\frac{3}{2}}\right\rangle$&$-2f_{\beta}$&$-\frac{17\sqrt{3}}{45}g_A$\\
			&\multirow{2}{*}{$P_{\psi_{ss}}^{N^0}$}&$\left\langle \Omega_cD, ^{2}S_{\frac{1}{2}}\right|\mathscr{L}\left|\Omega_cD^*, ^{4}S_{\frac{3}{2}}\right\rangle$&$-g_{\alpha}+f_{\alpha}$&$-\frac{2}{5}g_A$\\
			&&$\left\langle \Omega_cD^*, ^{2}S_{\frac{1}{2}}\right|\mathscr{L}\left|\Omega_cD^*, ^{4}S_{\frac{3}{2}}\right\rangle$&$-g_{\beta}+f_{\beta}$&$-\frac{2\sqrt{3}}{45}g_A$\\
			&\multirow{2}{*}{$P_{\psi_{ss}}^{N^-}$}&$\left\langle \Omega_cD, ^{2}S_{\frac{1}{2}}\right|\mathscr{L}\left|\Omega_cD^*, ^{4}S_{\frac{3}{2}}\right\rangle$&$g_{\alpha}-f_{\alpha}$&$\frac{2}{5}g_A$\\
			&&$\left\langle \Omega_cD^*, ^{2}S_{\frac{1}{2}}\right|\mathscr{L}\left|\Omega_cD^*, ^{4}S_{\frac{3}{2}}\right\rangle$&$g_{\beta}-f_{\beta}$&$\frac{2\sqrt{3}}{45}g_A$\\
			\bottomrule[1.0pt]
			\bottomrule[1.0pt]
	\end{tabular}
\end{table*}
\renewcommand\tabcolsep{0.00cm}
\renewcommand{\arraystretch}{1.80}
\begin{table*}[!htbp]
	\caption{The axial-vector coupling constants of the octet hidden-charm pentaquark states with the $8_{2f}$ flavor representation.}
	\label{TableAx2}
		\begin{tabular}{c|c|c|c|c}
			\toprule[1.0pt]
			\toprule[1.0pt]
			Processes&&&Coefficients&Results\\
			\hline
			\multirow{8}{*}{$\frac{1}{2}^-\to\frac{1}{2}^-$}&$P_{\psi}^{N^+}$&$\left\langle\Lambda_cD, ^{2}S_{\frac{1}{2}}\right|\mathscr{L}\left|\Lambda_cD^*, ^{2}S_{\frac{1}{2}}\right\rangle$&$g_{\kappa}+f_{\kappa}$&$-\frac{\sqrt{3}}{5}g_A$\\
			&$P_{\psi}^{N^0}$&$\left\langle\Lambda_cD, ^{2}S_{\frac{1}{2}}\right|\mathscr{L}\left|\Lambda_cD^*, ^{2}S_{\frac{1}{2}}\right\rangle$&$-g_{\kappa}-f_{\kappa}$&$\frac{\sqrt{3}}{5}g_A$\\
			&$P_{\psi_s}^{\Sigma^+}$&$\left\langle\Xi_cD, ^{2}S_{\frac{1}{2}}\right|\mathscr{L}\left|\Xi_cD^*, ^{2}S_{\frac{1}{2}}\right\rangle$&$2f_{\kappa}$&$-\frac{\sqrt{3}}{5}g_A$\\
			&$P_{\psi_s}^{\Sigma^0}$&$\left\langle\Xi_cD, ^{2}S_{\frac{1}{2}}\right|\mathscr{L}\left|\Xi_cD^*, ^{2}S_{\frac{1}{2}}\right\rangle$&$\frac{2}{\sqrt{3}}g_{\kappa}$&$-\frac{1}{5}g_A$\\
			&$P_{\psi_s}^{\Lambda^0}$&$\left\langle\Lambda_cD, ^{2}S_{\frac{1}{2}}\right|\mathscr{L}\left|\Lambda_cD^*, ^{2}S_{\frac{1}{2}}\right\rangle$&$\frac{2}{\sqrt{3}}g_{\kappa}$&$-\frac{1}{5}g_A$\\
			&$P_{\psi_s}^{\Sigma^-}$&$\left\langle\Xi_cD, ^{2}S_{\frac{1}{2}}\right|\mathscr{L}\left|\Xi_cD^*, ^{2}S_{\frac{1}{2}}\right\rangle$&$-2f_{\kappa}$&$\frac{\sqrt{3}}{5}g_A$\\
			&$P_{\psi_{ss}}^{N^0}$&$\left\langle\Xi_cD, ^{2}S_{\frac{1}{2}}\right|\mathscr{L}\left|\Xi_cD^*, ^{2}S_{\frac{1}{2}}\right\rangle$&$-g_{\kappa}+f_{\kappa}$&0\\
			&$P_{\psi_{ss}}^{N^-}$&$\left\langle\Xi_cD, ^{2}S_{\frac{1}{2}}\right|\mathscr{L}\left|\Xi_cD^*, ^{2}S_{\frac{1}{2}}\right\rangle$&$g_{\kappa}-f_{\kappa}$&0\\
			\hline
			Processes&&&Coefficients&Results\\
			\hline
			\multirow{16}{*}{$\frac{3}{2}^-\to\frac{1}{2}^-$}&\multirow{2}{*}{$P_{\psi}^{N^+}$}&$\left\langle\Lambda_cD, ^{2}S_{\frac{1}{2}}\right|\mathscr{L}\left|\Lambda_cD^*, ^{4}S_{\frac{3}{2}}\right\rangle$&$g_{\tau}+f_{\tau}$&$-\frac{3}{5}g_A$\\
			&&$\left\langle\Lambda_cD^*, ^{2}S_{\frac{1}{2}}\right|\mathscr{L}\left|\Lambda_cD^*, ^{4}S_{\frac{3}{2}}\right\rangle$&$g_{\omega}+f_{\omega}$&$-\frac{\sqrt{3}}{5}g_A$\\
			&\multirow{2}{*}{$P_{\psi}^{N^0}$}&$\left\langle\Lambda_cD, ^{2}S_{\frac{1}{2}}\right|\mathscr{L}\left|\Lambda_cD^*, ^{4}S_{\frac{3}{2}}\right\rangle$&$-g_{\tau}-f_{\tau}$&$\frac{3}{5}g_A$\\
			&&$\left\langle\Lambda_cD^*, ^{2}S_{\frac{1}{2}}\right|\mathscr{L}\left|\Lambda_cD^*, ^{4}S_{\frac{3}{2}}\right\rangle$&$-g_{\omega}-f_{\omega}$&$\frac{\sqrt{3}}{5}g_A$\\
			&\multirow{2}{*}{$P_{\psi_s}^{\Sigma^+}$}&$\left\langle\Xi_cD, ^{2}S_{\frac{1}{2}}\right|\mathscr{L}\left|\Xi_cD^*, ^{4}S_{\frac{3}{2}}\right\rangle$&$2f_{\tau}$&$-\frac{3}{5}g_A$\\
			&&$\left\langle\Xi_cD^*, ^{2}S_{\frac{1}{2}}\right|\mathscr{L}\left|\Xi_cD^*, ^{4}S_{\frac{3}{2}}\right\rangle$&$2f_{\omega}$&$-\frac{\sqrt{3}}{5}g_A$\\
			&\multirow{2}{*}{$P_{\psi_s}^{\Sigma^0}$}&$\left\langle\Xi_cD, ^{2}S_{\frac{1}{2}}\right|\mathscr{L}\left|\Xi_cD^*, ^{4}S_{\frac{3}{2}}\right\rangle$&$\frac{2}{\sqrt{3}}g_{\tau}$&$-\frac{\sqrt{3}}{5}g_A$\\
			&&$\left\langle\Xi_cD^*, ^{2}S_{\frac{1}{2}}\right|\mathscr{L}\left|\Xi_cD^*, ^{4}S_{\frac{3}{2}}\right\rangle$&$\frac{2}{\sqrt{3}}g_{\omega}$&$-\frac{1}{5}g_A$\\
			&\multirow{2}{*}{$P_{\psi_s}^{\Lambda^0}$}&$\left\langle\Lambda_cD, ^{2}S_{\frac{1}{2}}\right|\mathscr{L}\left|\Lambda_cD^*, ^{4}S_{\frac{3}{2}}\right\rangle$&$\frac{2}{\sqrt{3}}g_{\tau}$&$-\frac{\sqrt{3}}{5}g_A$\\
			&&$\left\langle\Lambda_cD^*, ^{2}S_{\frac{1}{2}}\right|\mathscr{L}\left|\Lambda_cD^*, ^{4}S_{\frac{3}{2}}\right\rangle$&$\frac{2}{\sqrt{3}}g_{\omega}$&$-\frac{1}{5}g_A$\\
			&\multirow{2}{*}{$P_{\psi_s}^{\Sigma^-}$}&$\left\langle\Xi_cD, ^{2}S_{\frac{1}{2}}\right|\mathscr{L}\left|\Xi_cD^*, ^{4}S_{\frac{3}{2}}\right\rangle$&$-2f_{\tau}$&$\frac{3}{5}g_A$\\
			&&$\left\langle\Xi_cD^*, ^{2}S_{\frac{1}{2}}\right|\mathscr{L}\left|\Xi_cD^*, ^{4}S_{\frac{3}{2}}\right\rangle$&$-2f_{\omega}$&$\frac{\sqrt{3}}{5}g_A$\\
			&\multirow{2}{*}{$P_{\psi_{ss}}^{N^0}$}&$\left\langle\Xi_cD, ^{2}S_{\frac{1}{2}}\right|\mathscr{L}\left|\Xi_cD^*, ^{4}S_{\frac{3}{2}}\right\rangle$&$-g_{\tau}+f_{\tau}$&0\\
			&&$\left\langle\Xi_cD^*, ^{2}S_{\frac{1}{2}}\right|\mathscr{L}\left|\Xi_cD^*, ^{4}S_{\frac{3}{2}}\right\rangle$&$-g_{\omega}+f_{\omega}$&0\\
			&\multirow{2}{*}{$P_{\psi_{ss}}^{N^-}$}&$\left\langle\Xi_cD, ^{2}S_{\frac{1}{2}}\right|\mathscr{L}\left|\Xi_cD^*, ^{4}S_{\frac{3}{2}}\right\rangle$&$g_{\tau}-f_{\tau}$&0\\
			&&$\left\langle\Xi_cD^*, ^{2}S_{\frac{1}{2}}\right|\mathscr{L}\left|\Xi_cD^*, ^{4}S_{\frac{3}{2}}\right\rangle$&$g_{\omega}-f_{\omega}$&0\\
			\bottomrule[1.0pt]
			\bottomrule[1.0pt]
	\end{tabular}
\end{table*}

\section{Summary}\label{sec5}
In recent years, the researchs about hidden-charmed pentaquark states had great progress, most of the studies were focused on the properties of pentaquark states such as mass and decay width. Inspired by these studies, we systematically investigated the radiative decay and axial-vector decay behaviors of octet hidden-charmed pentaquark states.

In this work, we construct the wave functions of the octet hidden-charm pentaquark states with constituent quark model, we calculate the transition magnetic moments and radiative decay widths of the octet pentaquark states. The numerical results show that the transition magnetic moments and radiative decay widths of pentaquark states with the same flavor representations satisfy some relations. For example, for the $P_{\psi}^{N^+}$ state with $8_{1f}$ flavor representation, their transition magnetic moments satisfy the relation $\frac{\mu_{\Sigma_cD^*|\frac{3}{2}^-\rangle\to \Sigma_cD|\frac{1}{2}^-\rangle\gamma}}{\mu_{\Sigma_cD^*|\frac{1}{2}^-\rangle\to \Sigma_cD|\frac{1}{2}^-\rangle\gamma}}$=$\sqrt{2}$, their radiative decay widths $\Gamma_{\Sigma_cD^*|\frac{1}{2}^-\rangle\to \Sigma_cD|\frac{1}{2}^-\rangle\gamma}$ and $\Gamma_{\Sigma_cD^*|\frac{3}{2}^-\rangle\to \Sigma_cD|\frac{1}{2}^-\rangle\gamma}$ are quite close, and the decay width $\Gamma_{\Sigma_cD^*|\frac{3}{2}^-\rangle\to \Sigma_cD^*|\frac{1}{2}^-\rangle\gamma}$ is close to 0. We also calculated the axial-vector coupling constants for the transition processes in the chiral quark model, we notice that the $\pi$ meson interaction only exists among light quarks, and the axial-vector coupling constants are quite sensitive to the flavor and spin configurations. Among the pentaquark states we discussed, for the $P_{\psi_{ss}}^{N^0}$ and $P_{\psi_{ss}}^{N^-}$ states with $8_{2f}$ flavor representation, the axial-vector coupling constants of the processes $P_{\psi}|\frac{1}{2}^-\rangle\to P_{\psi}|\frac{1}{2}^-\rangle\pi_0$, $P_{\psi}|\frac{3}{2}^-\rangle\to P_{\psi}|\frac{1}{2}^-\rangle\pi_0$ are all 0.

The transition magnetic moments and decay widths play important role in the studies of inner structure of pentaquark states, and can also distinguish experimentally between pentaquark states with different configurations. The axial-vector coupling constants can help us understand the strong interactions, and are also helpful for the calculations of chiral effective theory. We hope that the present work will inspire more exploration of the properties related to octet hidden-charmed pentaquark states.

\section*{Acknowledgement}
This project is supported by the National Natural Science Foundation of China under Grants No. 11905171. This work is also supported by Shaanxi Fundamental Science Research Project for Mathematics and Physics (Grant No. 22JSQ016) and Young Talent Fund of Xi'an Association for Science and Technology (Grant No. 959202413087).

\end{document}